\documentclass[lettersize,journal]{IEEEtran}

\usepackage[table, dvipsnames]{xcolor}
\usepackage{amsmath,amsfonts}
\usepackage{algorithmic}
\usepackage{algorithm}
\usepackage{glossaries}
\usepackage{longtable}
\usepackage{glossary-mcols}
\usepackage{array}
\usepackage{pgfplots}
\usepackage{multirow}
\usepackage[caption=false,font=normalsize,labelfont=sf,textfont=sf]{subfig}
\usepackage{textcomp}
\usepackage{stfloats}
\usepackage{url}
\usepackage[inline]{enumitem}
\usepackage{verbatim}
\usepackage{tabularx}
\usepackage{amssymb}
\usepackage{booktabs}
\usepackage{cite}
\usepackage{pgfplots}
\usepackage{soul}
\usepackage{hyperref}
\usepackage{graphicx}

\makeglossaries 
\newacronym{eo}{EO}{Earth Observation}
\newacronym{firms}{FIRMS}{Fire Information for Resource Management System}
\newacronym{fso}{FSO}{Free Space Optics}
\newacronym{haps}{HAPS}{High-Altitude Platform Stations}
\newacronym{hapgs}{HAPGS}{High-Altitude Platform Ground Stations}
\newacronym{hale}{HALE}{High-Altitude Long-Endurance}
\newacronym{hetnet}{HetNet}{Heterogeneous Network}
\newacronym{isl}{ISL}{{Inter-Satellite Links}}
\newacronym{laps}{LAPS}{Low-Altitude Platforms}
\newacronym{lapse}{LAPSE}{Low-Altitude Platform Eavesdropper}
\newacronym{leo}{LEO}{Low Earth Orbit}
\newacronym{leoe}{LEOE}{Low Earth Orbit Eavesdropper}
\newacronym{leom}{LEOM}{Low Earth Orbit Mothership}
\newacronym{modis}{MODIS}{Moderate Resolution Imaging Spectroradiometer}
\newacronym{nasa}{NASA}{{National Aeronautics and Space Administration}}
\newacronym{nfpa}{NFPA}{National Fire Protection Association}
\newacronym{rf}{RF}{Radio Frequency}
\newacronym{snr}{SNR}{Signal-to-Noise Ratio}
\newacronym{sos}{SoS}{System-of-Systems}
\newacronym{sparta}{SPARTA}{{Space Attack Research and Tactic Analysis}}
\newacronym{ttc}{TT\&C}{Telemetry, Tracking, and Command}
\newacronym{uavs}{UAVs}{Unmanned Aerial Vehicles}
\newacronym{viirs}{VIIRS}{Visible Infrared Imaging Radiometer Suite}

\glsaddall[types=\acronymtype]

\newglossarystyle{customacronyms}
{
    \setglossarystyle{long3col}

}

\begin{document}

\title{Securing {Heterogeneous Network (HetNet)} Communications for Wildfire Management: Mitigating the Effects of Adversarial and Environmental Threats}

\author{Nesrine~Benchoubane,~\IEEEmembership{Graduate Student Member,~IEEE,} Olfa~Ben~Yahia,~\IEEEmembership{Member,~IEEE,} William~Ferguson,~\IEEEmembership{Member,~IEEE,} Gürkan~Gür,~\IEEEmembership{Senior Member,~IEEE,} Sumit~Chakravarty,~\IEEEmembership{Member,~IEEE,} 
Gregory~Falco,~\IEEEmembership{Member,~IEEE}
and~Gunes~Karabulut~Kurt,~\IEEEmembership{Senior Member,~IEEE,}
\thanks{Nesrine~Benchoubane, Olfa~Ben~Yahia, and Gunes~Karabulut~Kurt are with the Poly-Grames Research Center, Department of Electrical Engineering, Polytechnique Montréal, Montreal, QC H3C 3A7, Canada (e-mail: nesrine.benchoubane@polymtl.ca; olfa.ben-yahia@polymtl.ca; gunes.kurt@polymtl.ca).}
\thanks{William Ferguson is with the 303 Overwatch (A.S.B.L), Luxembourg City, Luxembourg  (e-mail: william.o.ferguson@ethicallyhacking.space).}
\thanks{Gürkan Gür is with the Institute of Computer Science (InIT), Zurich University of Applied Sciences (ZHAW), 8401 Winterthur, Switzerland (email: gurkan.gur@zhaw.ch).}
\thanks{Sumit Chakravarty is with the Department of Electrical and Computer Engineering, Kennesaw State University, Kennesaw, GA 30144, United States 
(e-mail: schakra2@kennesaw.edu).}
\thanks{Gregory Falco is with the Sibley School of Mechanical and Aerospace Engineering, Cornell University, Ithaca, NY 14853,  United States 
(e-mail: gfalco@cornell.edu).}
}
\markboth{IEEE JOURNAL OF RADIO FREQUENCY IDENTIFICATION, VOL. X, 2024}
{BENCHOUBANE \MakeLowercase{\textit{et al.}}: A Sample Article Using IEEEtran.cls for IEEE Journals}


\maketitle

\begin{abstract}
In the face of adverse environmental conditions and cyber threats, robust communication systems for critical applications such as wildfire management and detection demand secure and resilient architectures.  This paper presents a novel framework that considers both {adversarial factors}, building resilience into a heterogeneous network {(HetNet)} integrating \Gls{leo} satellite  constellation with \Gls{hapgs} and \Gls{laps}{, tailored to support wildfire management operations}. Building upon our previous work on \textit{secure-by-component} approach for link segment security, we extend protection to the communication layer by securing both \Gls{rf}/\Gls{fso} management {and different links}. {Through a case study, we quantify how environmental stressors impact secrecy capacity and expose the system to passive adversaries. Key findings demonstrate that atmospheric attenuation and beam misalignment can notably degrade secrecy capacity across both short- and long-range communication links, while high-altitude eavesdroppers face less signal degradation, increasing their interception capability. Moreover, increasing transmit power to counter environmental losses can inadvertently improve eavesdropper reception, thereby reducing overall link confidentiality. Our work} not only highlights the importance of protecting networks from these dual threats but also aligns with the IEEE P3536 Standard for Space System Cybersecurity Design, ensuring resilience and the prevention of mission failures. 
\end{abstract}

\begin{IEEEkeywords}
Earth observation (EO), HetNet, hybrid RF/FSO, link security, space cybersecurity.
\end{IEEEkeywords}

\newpage

\section{Introduction} \label{sec:intro}

\IEEEPARstart{W}{ildfires} — also known as bushfires or forest fires — are uncontrolled blazes that spread across forests, shrublands, grasslands, savannas, and croplands \cite{owid-wildfires}. It is a growing global threat, exacerbated by climate change, with devastating impacts on ecosystems and the global economy. {It} can spread uncontrollably, consuming vast areas of land and threatening biodiversity. This is exemplified by historical events such as Australia’s 2019–2020 Black Summer that burned over {10 million hectares\cite{NSW2021Fire}}. Additionally, {as reported by} \cite{Danielle2025LAWildfires}, the economic toll has experienced a sharp rise. It is estimated that wildfires in Los Angeles alone caused over \$250 billion in damages and economic losses in January 2025, surpassing the total losses from the entire 2020 wildfire season. Beyond financial losses, wildfires also contribute significantly to global carbon emissions. Each year, wildfires release an estimated 5 to 8 billion tonnes of $\mathrm{CO}_\mathrm{2}$ into the atmosphere, exacerbating climate change \cite{owid-wildfires}. Another consequence is the increased frequency of extreme weather events and disruption of agricultural patterns and water resources, {leading to} socioeconomic challenges. 

{While} natural factors such as lightning strikes do contribute to wildfires, the overwhelming majority are human-induced. Arson, land-use changes, agricultural practices, and carelessness during outdoor activities account for over 85\% of all wildfire \cite{doi:10.1073/pnas.1617394114}. In particular, arson has become a significant cause of deliberate wildfire outbreaks, presenting unique challenges for wildfire detection and management. Arsonists often target remote, difficult-to-monitor areas, intentionally igniting fires that can spread quickly and devastate large regions. For instance, between 2014 and 2018, \Gls{nfpa} recorded 52,260 intentional fire incidents in the U.S., resulting in 400 civilian deaths, 950 injuries, and \$815 million in annual property {damage \cite{Campbell2021IntentionalFires, Haynes_2019}.} The threat extends to first responders who, while working to control the fires, also risk their lives in these hazardous environments. 

These incidents underscore the need for advanced technologies capable of managing and detecting both accidental and deliberate fire ignitions in real-time. Military and defense organizations are particularly concerned with the threat posed by arson in strategic areas, such as near military bases or key infrastructure. As such, improving the management and detection of early fire signatures and integrating security measures into communication networks {becomes crucial for national security}. From a cybersecurity {and resilience} standpoint, adversarial actors can exploit the vulnerabilities in communication and detection systems, triggering delays or misdirecting resources, further complicating wildfire management efforts. 

\subsection{Wildfire Detection and Management Systems}

Over the last two decades, satellite-based \Gls{eo} systems have revolutionized the detection and monitoring of wildfires. These systems, such as {the \Gls{nasa}}’s \Gls{firms}, have provided real-time active fire data for over a decade, using instruments such as the \Gls{modis} and the \Gls{viirs} \cite{Blumenfeld2020Wildfires}. FIRMS can detect hotspots within three hours of satellite observation and provide imagery within four to five hours, improving wildfire response times. Thermal infrared observations from satellites are increasingly seen as a cost-effective way to detect wildfires over large areas, but despite decades of active fire mapping, {current fire detection systems continue to face persistent challenges in communications and surveillance~\cite{KopardekarGrindle2021}}. Additionally, traditional satellite imaging is often limited to only four or five snapshots per day, leaving critical gaps in fire tracking. This becomes even more critical when attempting to detect and combat wildfires started by malicious actors, such as arsonists, who strategically set fires in areas that are difficult to monitor.

To address these gaps, the Earth Fire Alliance plans to deploy FireSat, a first-of-its-kind satellite constellation dedicated to global wildfire monitoring, with its protoflight launched in March 2025. The full constellation of 50+ satellites is expected by 2030 \cite{McHugh-Johnson2025FireSat}. Similarly, Canada’s WildFireSat \cite{CSA2025WildFireSat}, a planned constellation of seven microsatellites, is set to launch in 2029, aiming to provide wildfire managers with near-real-time {coverage}. However, even these systems have limitations. Satellite systems may fail to detect small, localized fires in the critical early stages before they spread uncontrollably. In that effect, \Gls{haps} and \Gls{hale} {vehicles have been seen} as a promising solution for persistent fire monitoring. In 2019, NASA initiated a program to explore at-altitude concepts for wildfire surveillance \cite{6GWorld2025HAPS}, and a study in \cite{Burns2023HAPS} has estimated that monitoring a single fire season would require either eight \Gls{haps} vehicles or 23 fixed-wing aircraft. {However, these systems rely} on favorable weather conditions which makes them susceptible to operational disruptions. 

\subsection{Emerging Architectures}

While these technologies provide significant improvements in wildfire management and detection, their full potential can be realized by integrating them into a cohesive communication and data-sharing framework. Emerging architectures such as \Gls{hetnet} offer a promising solution by interconnecting multiple platforms, {including satellites and HAPS}, to provide continuous and resilient coverage \cite{10589561, 8030545}. {Unlike satellite-only or aerial-only systems—which face limitations in coverage granularity, latency, or susceptibility to environmental conditions \cite{KopardekarGrindle2021}— as mentioned previously, HetNet} leverage multi-layered network architectures {that combine both system types}. {Additionally, it integrates dual} \Gls{rf} and \Gls{fso} communication to facilitate low-latency, high-bandwidth data transmission between spaceborne and airborne assets \cite{9129403, 10570308}. {This dual RF/FSO enhances connectivity and} enables adaptive networking, where data routing is optimized dynamically based on environmental conditions and system constraints \cite{9655260}{, and channel allocation optimization is used \cite{9832657, 10185601}.}

{Additionally, the importance of maintaining robust communication links becomes evident in large-scale incidents such as the Australian Black Summer fires, where communication breakdowns disrupted coordination efforts \cite{NSW2021Fire}. The disruptions underscore the operational need for resilient and adaptive communication infrastructures which can be provided by the HetNet architectures}. We emphasize this adaptability as it is particularly critical in our context {as it} can severely hinder firefighting efforts. We also stress how in a {homeland security} context, these networks would provide the added benefit of enabling continuous situational awareness of areas at risk of wildfire, as well as enhancing the coordination of defense resources in the event of an intentional fire or an attack on critical infrastructure. 

\subsection{Cyber Adversary}

In the design of this HetNet, we prioritize the protection of communication links and network components against potential cyber threats. As these networks rely heavily on space-based and airborne assets, securing the communication infrastructure becomes critical, particularly in environments where disruption could have severe {consequences. Existing research} \cite{10850060} analyzes attack surfaces in optical links and applies a \textit{secure-by-component} methodology to the link segment. Other works \cite{10539344, 10615868} propose secure multi-layer airborne FSO backhaul networks and \Gls{hapgs} for optical \Gls{leo} satellite constellations, enhancing data transmission resilience, while \cite{doi:10.2514/6.2023-4800} present a comprehensive attack tree analysis of HAPS. Additionally, physical layer security strategies have been proposed for hybrid \Gls{fso}/\Gls{rf} networks \cite{MOHSAN2023100697, 9691902, 9655260, 9806158}, which include dynamic link selection based on weather conditions and secrecy metric analysis \cite{9238951}. Thus, we emphasize the importance of incorporating protection measures that not only defend against cyber threats but also ensure the reliability of {communication links}. 

\begin{figure*}[tb]
    \centering
    \includegraphics[width=\linewidth]{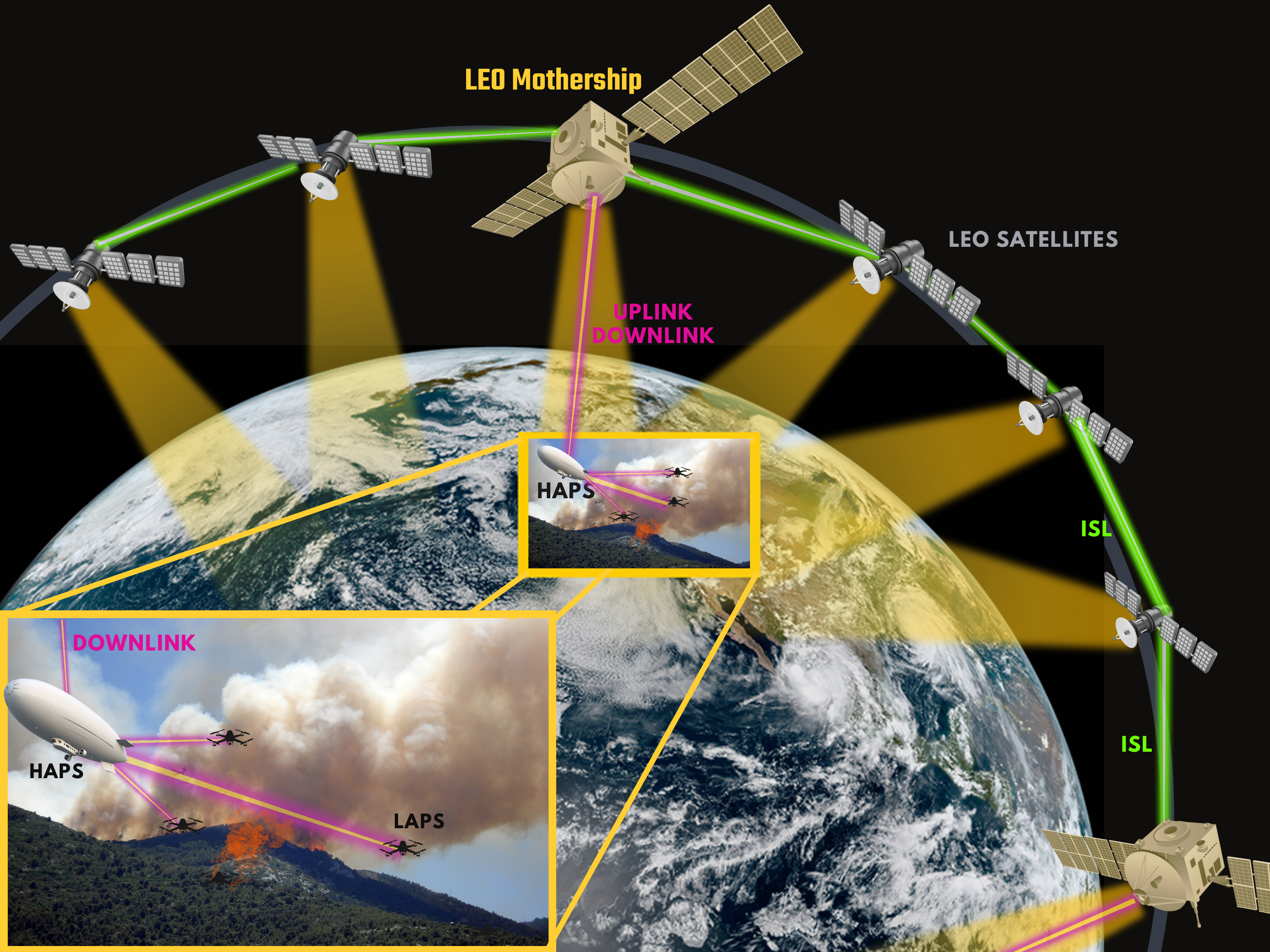}
    \caption{{Proposed wildfire management SoS architecture, showcasing: (1) the LEO constellation, where satellites perform EO observations and communicate via {\Gls{isl}}; (2) the \Gls{leom} and \Gls{hapgs}, which establish downlink and uplink connections for data relay; and (3) the \Gls{hapgs} deployed over wildfire regions, coordinating with LAPS that operate at lower altitudes with established downlinks for real-time tactical monitoring.}}
\label{fig:wildfire-solution}
\end{figure*}

\subsection{A Novel Perspective: Nature is also an Adversary}

The novelty of our paper lies in broadening the adversarial perspective to include both cyber threats and environmental disturbances, treating nature itself as a potential adversary. This work significantly extends \cite{10850060}, which only considered a single layer and focused solely on cyber threats. To the best of our knowledge, we are also the first to incorporate this dual perspective into cybersecurity modeling {of these systems}. Unlike traditional approaches that primarily focus on malicious human actors, our model also takes into account how environmental conditions, such as wildfire-induced disruptions, can compromise communication reliability. These disturbances can provide adversaries with opportunities to exploit vulnerabilities in communication links, especially during periods of heightened risk, such as when FSO and RF signals are disrupted by smoke plumes. {By emphasizing the interplay between cyber and environmental threats, our work addresses a critical research gap: existing models often treat these risks separately, overlooking their combined potential to amplify vulnerabilities. To address this shortcoming, our objective is to develop a secure and resilient HetNet communication framework that models this interplay as a unified and strategic threat surface, focusing on supporting wildfire management operations.}

\subsection{Research Contributions and Scope}

The main contributions of this paper are:
\begin{itemize}
    \item We propose a HetNet architecture that integrates future EO constellations with HAGS to support robust wildfire detection in extreme environments. This multi-tiered system ensures continuous Earth observation, leveraging diverse platforms to enhance monitoring coverage and responsiveness.
    \item {We align our approach with \textit{IEEE P3536 Standard for Space System Cybersecurity Design} \cite{IEEE2025P3536} process} to secure the design of FSO \& RF management, and Space-to-X links. This guarantees compliance with emerging cybersecurity frameworks, enhancing the resilience of communication links in dynamic and adversarial settings.
    \item We quantify the impact of environmental stressors on secrecy capacity and the trade-offs between resilience and security, analyzing atmospheric attenuation, eavesdropper power and distance, and transmit power through quantitative modeling and simulations.
    \item We develop novel protection techniques taking into account the cyber and harsh environment adversaries. These techniques mitigate both adversaries, ensuring reliable and secure data transmission across all network layers.
\end{itemize}

\subsection{Structure}

The rest of the paper is organized as follows. Section~\ref{architecture} describes the {network architecture}. Section~\ref{sec-by-design} presents the system design considerations. Section~\ref{sec:threat model} defines the adversary models for the system boundary considered. Section~\ref{case-study} delves deeper into the secrecy performance in such circumstances and provides a concrete analysis regarding the eavesdropping attack under environmental disturbances. Then, building on the findings of previous sections, Section~\ref{sec:protect} defines the protection techniques to be designed in the system to maintain secure, reliable operations. {Section~\ref{takeways} presents the key takeaways and limitations.} Section~\ref{conclusion} concludes the paper.

\section{Wildfire Management System of System}
\label{architecture}

The proposed wildfire management network is designed as a \Gls{sos} architecture that integrates a variety of heterogeneous elements across multiple tiers, each performing distinct functions. {We highlight the different network layers and communication links in Fig.~\ref{fig:wildfire-solution}, which we detail further below.}

\subsection{Functional Layers and System Segments}

The network is structured into multiple functional layers, where each layer consists of nodes responsible for specific tasks related to data acquisition, coordination, and decision-making. 
\begin{enumerate}[label=(\roman*)]
 \item \textbf{LEO Satellites (500 to 2000 km) - \textit{EO Data Generation and Relay}}

        LEO satellites serve as the primary EO nodes in the network, providing continuous surveillance and data relay capabilities. They are equipped with thermal imaging, multispectral, and SAR payloads to detect fire outbreaks and monitor their progression over time. A subset of these nodes serves as mothership nodes (M), denoted LEOM, responsible for coordinating observations and data aggregation from other LEO nodes and relaying critical data to the HAPS nodes.
        \item \textbf{HAPS (40 km) – \textit{Coordination and Decision-Making}}
        
        A HAPS node acts as a high-altitude ground station in the stratosphere, denoted HAPGS, acting also as a relay between LEO satellites and lower-altitude network components. Their primary function is to process EO data received from LEO satellites and distribute operational commands to LAPS based on real-time fire assessments. They also coordinate safe operational zones for LAP deployment and optimize sensor coverage over affected areas.
        \item \textbf{LAPS (1 to 5 km) – \textit{Low-Altitude Tactical Monitoring and Sensor Coordination}}
        
        LAPS including \Gls{uavs} and drones operate at low altitudes, providing high-resolution, real-time data in wildfire-affected regions. These platforms are equipped with infrared cameras, gas sensors, and environmental detectors to measure fire intensity, smoke dispersion, and air quality. LAPS are dynamically deployed by HAPS and can autonomously adjust their flight paths to maximize data collection efficiency. 
    \end{enumerate}

In the SoS, we can also identify the different segments related to the functional layers of the architecture:  \begin{enumerate*}
    \item \textbf{Space Segment} which includes the LEO constellation that enables data collection on-orbit and relay capabilities from the LEOM back to the ground. \item \textbf{Ground segment} which includes all the HAPGS nodes in the system that manage communication, data processing, and relay to the user. \item \textbf{User Segment} which consists of the LAPS nodes that support the execution of tasks in the field. \item \textbf{Integration segment}, which is responsible for enabling the integration and interoperability between the various segments of the SoS. \item \textbf{Link segment} which encompasses all the communication pathways that connect the segments.
\end{enumerate*}

\subsection{The Crucial Enabler: Link Segment}

\begin{table*}[t!]
    \centering
    \renewcommand{\arraystretch}{1.8} 
    \caption{TT\&C communication links overview.}
    \begin{tabular}{|l|l|p{10cm}|}
        \hline
        \textbf{Direction} & \textbf{TT\&C Function} & \textbf{Description} \\
        \hline \hline
        \multirow{3}{*}{Uplink (HAPGS $\rightarrow$ LEOM)}  
            & Command (C - $\clubsuit$) & Issues mission directives for EO data collection and constellation coordination. \\
            & Telemetry (T - $\diamondsuit$) & Sends HAPGS-level network diagnostics, link stability, and environmental conditions. \\
            & Tracking (T - $\triangledown$) & Maintains FSO beam alignment and compensates for LEOM motion. \\
        \hline
        \multirow{3}{*}{ISL (LEOM $\leftrightarrow$ LEO Constellation)}  
            & Telemetry (T - $\diamondsuit$) & Tracks ISL performance, satellite health, and relayed data integrity. \\
            & Tracking (T - $\triangledown$) & Synchronizes satellite positioning for stable ISL communication. \\
        \hline
        \multirow{3}{*}{Downlink (LEOM $\rightarrow$ HAPGS)}  
            & Telemetry (T - $\diamondsuit$) & Monitors LEOM status and downlinks aggregated EO data to HAPGS. \\
            & Tracking (T - $\triangledown$) & Ensures continuous FSO beam tracking and handover between satellites and HAPGS. \\
        \hline
        \multirow{3}{*}{Downlink (HAPGS $\rightarrow$ LAPS)}  
            &  Command (C - $\clubsuit$) & Issues tasking commands for tactical wildfire monitoring operations. \\
            & Telemetry (T - $\diamondsuit$) & Provides LAPS status updates and sensor performance data. \\
            & Tracking (T - $\triangledown$) & Maintains stable data transmission to LAPS under dynamic environmental conditions. \\
        \hline
    \end{tabular}
    \label{tab:ttc_links}
\end{table*}

The link segment plays a pivotal role in ensuring reliable and uninterrupted communication between the various components of the SoS architecture. Therefore, we provide a detailed analysis on the impact of adversaries on communication links in Section~\ref{sec:threat model} while putting this effort in the context of secure system engineering compliant to the \textit{IEEE P3536 Standard for Space System Cybersecurity Design} in the next section.

In our HetNet, the nodes within the network employ a hybrid FSO/RF communication to maintain {robust} data transmission across different segments, which is crucial for ensuring the system’s resilience to environmental challenges and adversarial interference. In addition, \Gls{ttc} functions are integral to the operational integrity of the network, maintaining link stability and executing mission directives. {Table~\ref{tab:ttc_links} summarizes the communication links, associated TT\&C functions, and their roles. For clarity, we denote command links as $\clubsuit$, telemetry as $\diamondsuit$, and tracking as $\triangledown$.}

\section{System Design Compliance with IEEE P3536 Standard}
\label{sec-by-design}

 Given the complexity of the operational environment, the design of the SoS in all its layers must address both environmental and cyber adversaries while maintaining robust communication and integration across the space, ground, and user segments. Thus, it must be engineered for resilience and security, and the communication infrastructure, defined within the link segment, is particularly critical, as it underpins system operations and is highly susceptible to failure due to both adversaries. 
 
 Ensuring system-wide compliance with the \textit{IEEE P3536 Standard for Space System Cybersecurity Design} \cite{IEEE2025P3536} provides a structured framework to mitigate mission failure. The standard has been demonstrated in \cite{10592289, 10795027} to prove that each segment—whether space-based, ground-based, or user-facing—must meet rigorous security and resilience criteria and follow a series of subprocesses that are designed to methodically address the system's engineering needs. 

 \vspace{0.2cm}
\noindent The subprocesses for system design are as follows: \begin{enumerate}[label=\textbf{\textsc{Subprocess \arabic*}}, left=0pt, labelsep=1em]
    \item Establish mission objective and failure modes of interest to technical system boundary.
    \item Establish threat techniques to attack surface.
    \item Establish protection techniques to technical traceability.
    \item Establish secure blocks.
\end{enumerate}

Each subprocess will be critical in aligning system components with operational and security goals. In the following sections, we will walk through each of these subprocesses in detail, providing insight into how they collectively contribute to a robust and secure system design. Appendix \ref{appendix1} consolidates all relevant details to facilitate a clearer understanding of the system's security design.

\vspace{0.3cm}

\noindent {\textbf{\textsc{Subprocess 1: Establish mission objective and failure modes of interest to technical system boundary}}}

\noindent The first subprocess involves defining mission objectives, identifying failure modes, and determining the system’s technical boundaries. 

\subsection{{Mission Objective}}

The wildfire management mission aims to enable real-time detection, monitoring, and coordinated response to protect lives, infrastructure, and ecosystems. Reliable, uninterrupted communication across all system layers is essential for situational awareness and rapid decision-making. Integrating aerial platforms with LEO satellites support enhances coverage while reducing risks to first responders. However, harsh environments introduce uncertainties that can impact system performance. 


\subsection{{Failure Modes}}

One of the most critical failure modes is the disruption of communication between HAPGS and the LEO constellation, which cascades into the loss of TT\&C. This failure severely impacts wildfire management operations, as it prevents real-time coordination and data relay essential for situational awareness. Additionally, the disruption of communication between the HAPGS and LAPS would further exacerbate operational challenges by isolating critical user nodes from the system. The underlying causes of these failures can be categorized into two primary domains:

\begin{itemize}
    \item \textit{Environmental Disturbances}: Wildfires generate extreme heat, dense smoke, and airborne particulates that interfere with FSO and RF communications and thus contribute to signal attenuation, beam misalignment, and increased background noise, impacting data transmission reliability. 
    \item \textit{Cyber Adversaries}: Threat actors may leverage the degraded environment to launch different attacks further increasing the risk of communication loss.
\end{itemize}

\subsection{{System Boundary}}

The system boundary is established by identifying critical mission elements susceptible to failure and defining their operational roles. Given that the two failure modes relate to communication loss, the system boundary includes technical elements responsible for maintaining data availability and transmission under extreme environmental and adversarial conditions. The two key elements of the link segment that address these concerns are as follows:

\begin{itemize}
\item \textit{FSO/RF Management}: This element ensures the reliability and stability of FSO and RF links across all layers. This includes managing link performance in normal and extreme conditions and ensuring continuous communication despite disruptions.

\item \textit{Space-to-X Links}: This element ensures the communication and coordination between the space-based and ground-based nodes and encompasses the communication links between various network layers, including the space-to-air link between the LEOM-HAPGS ($\diamondsuit$ \& $\triangledown$) and the HAPGS-to-LAP link ($\clubsuit$).
\end{itemize}

\section{Adversarial Threats to the System Boundary}
\label{sec:threat model}

Given the identified failure modes and the system boundary components (FSO/RF Management and Space-to-X Links), we present a comprehensive inventory of adversarial techniques that could lead to mission failure in this section.

\vspace{0.3cm}
\noindent \textbf{\textsc{Subprocess 2: Establish threat techniques to attack surface}}

\noindent The second subprocess involves mapping the attack surfaces of the elements in the system boundary to corresponding adversarial techniques across both {adversarial} domains.

\subsection{{Attack Surface}}

The attack surface is defined by vulnerabilities within the two technical elements of the system boundary as follows:

\begin{itemize}
\item \textit{FSO/RF Management}: This element is constrained by the limitations of adaptive link control, which struggles under dynamic interference caused by environmental disturbances such as smoke, low visibility, and particulate scattering. These disruptions impact beam alignment, signal coherence, and error correction, reducing the system's ability to maintain secure and continuous transmissions. Additionally, fallback mechanisms that switch between FSO and RF modes increase exposure to external influences, as adaptive thresholds may be exploited to degrade or manipulate link performance. 

\item \textit{Space-to-X Links}: This element operate across multiple transmission mediums, each with distinct exposure points. The LEOM-HAPGS and HAPGS-LAP links are vulnerable to environmental attenuation and interference, which may result in reduced signal clarity and an increased need for retransmissions, inadvertently extending the window of exposure. Additionally, the LEOs-LEOM links introduce additional risks on the LEOM node and can compromise the LEOM-HAPGS due to their reliance on coordinated network structures, where any LEO-compromised nodes or disruptions at key relay points could cascade across the system. 
\end{itemize}

\subsection{{Threat Techniques}}

\begin{table}[tpb!]
    \centering
    \renewcommand{\arraystretch}{1.8}
    \caption{{Mapping of passive threat techniques to attack surfaces.}}
    \begin{tabularx}{0.5\textwidth}{|X|X|X|}
        \hline
         \textbf{Domain} & \textbf{{Threat Technique}} & \textbf{Targeted Technical Element} \\
        \hline\hline
         Cyber Adversaries & {\textbf{Eavesdropping (\textit{EXF-0003})}} & Space-to-X Links \\
        Cyber Adversaries & Uplink Intercept (\textit{REC-0005.01}) & Space-to-X Links \\
         Cyber Adversaries & Downlink Intercept (\textit{REC-0005.02}) & Space-to-X Links \\
         Cyber Adversaries & Proximity Operations (\textit{REC-0005.03}) & Space-to-X Links \\
         Environmental Disruptions & Atmospheric Signal Attenuation (\textit{NAT-001}) & FSO/RF Management  \\
         Environmental Disruptions & Beam Misalignment  (\textit{NAT-002}) & FSO/RF Management \\
         Environmental Disruptions & Increased Background Noise (\textit{NAT-003}) & FSO/RF Management \\
        \hline
    \end{tabularx}
    \label{tab:sparta_attack_mapping}
\end{table}

{From the definition of the attack surfaces, we {identify} threat techniques arising from all {relevant adversaries. Specifically, cyber-related techniques are} assessed using the {\Gls{sparta}} framework \cite{aerospace2023sparta}. The full inventory is available in Appendix~\ref{appendix1}.}

{Among these, passive cyber techniques are of particular concern due to their unique overlap with environmental disturbances and their potential to facilitate larger, more disruptive attacks. These techniques exploit inherent weaknesses in communication channels, making them subtle yet highly impactful. A focused summary of these passive attacks is provided in Table~\ref{tab:sparta_attack_mapping} to relate the threat technique to the targeted technical element in the system. A prime example is eavesdropping (\textit{EXF-0003}), which we analyze in detail in the case study in the next section.}

\section{Case Study: Secrecy Performance Under Cyber \& Environmental Adversaries}
\label{case-study}

In this case study, we assess the combined impact of both environmental disturbances and cyber passive adversaries on communication link secrecy and overall network performance, which quantifies the effects it has on links of LEOM-HAPGS and HAPGS-LAP, assuming that both are FSO in this study, along with how they can be managed. We analyze various sub-techniques based on the mode of interception and the exploited communication medium, offering a comprehensive understanding of how eavesdropping can undermine network confidentiality and potentially lead to mission failure.

\subsection{Environmental Disturbances}

The LEOM-HAPGS and HAPGS-LAP FSO links, under various natural impairments, can be modeled to assess how these factors degrade the quality of communication as

\begin{equation}
    h_{main} = h_{att}  \cdot h_{pointing} \cdot h_{path-loss},
\end{equation}
where $ h_{att}$ is the attenuation factor due to atmospheric conditions. $ h_{pointing}$ is the pointing loss due to the misalignment of the optical beam. $ h_{path-loss}$ is the free-space path loss.

\subsubsection{Atmospheric Attenuation}

{This} refers to the reduction in signal strength as the optical signal travels through the atmosphere, {caused} by the smoke particles that scatter the light thus reducing the signal power as it propagates. The attenuation factor due to smoke and atmospheric conditions can be modeled as
\begin{equation}
    h_{att} = e^{-\alpha_{atm} \cdot d} \cdot G_t \cdot G_{r},
\end{equation}
where $\alpha_{atm}$ is the attenuation coefficient due to the atmosphere. $d$ is the propagation distance in the LEOM-HAPGS and HAPGS-LAP links. $G_t$ is the transmitter gain. $G_r$ is the receiver's gain. 

In the LEOM-HAPGS link, the transmitter gain $G_t$ is associated with LEOM, while in the HAPGS-LAP link, the transmitter gain $G_t$ is associated with HAPGS. Similarly, the receiver gain $G_r$ is associated with HAPGS in the LEOM-HAPGS link and with LAP in the HAPGS-LAP link.

According to Kim's model \cite{Kim2001ComparisonOL, Green2019}, {$\alpha_{atm}$ is influenced by factors such as the type of scattering, signal wavelength $ \lambda $, particle size distribution, and visibility}. The attenuation is calculated by
\begin{equation}
    \alpha_{atm} = f(\text{scattering type} , \text{particle size}, \lambda ,\text{visibility}),
    \label{eq}
\end{equation}
where {particle size distribution and scattering-type are determined by atmospheric conditions}. The smoke conditions in FSO in a controlled environment were reported in \cite{7000804}.

\subsubsection{Pointing Errors}

{This} occurs when the optical beam does not stay perfectly aligned between the transmitter and receiver, {which} can be modeled as

\begin{equation}
    h_{pointing} = \eta(\theta),
\end{equation}
where $\eta(\theta)$ is the pointing loss factor that depends on the pointing error angle $\theta$.

\subsubsection{Free-Space Path Loss}

{This} accounts for the geometrical spreading of the light beam as it travels over a distance {and} follows the inverse-square law, {which} can be modeled as

\begin{equation}
    h_{path-loss}= \Big( \frac{4 \pi d}{\lambda}\Big)^2,
\end{equation}
where $d$ is the distance in the LEOM-HAPGS and HAPGS-LAP {links.}

\noindent After considering all of these impairments, the received power $P_e$ at the receiver node can be expressed as

\begin{align}
P_{r} &= {P_t} G_t G_r \cdot e^{-\alpha_{atm} \cdot d} \cdot \eta(\theta) \cdot \frac{1}{\left(4\pi d / \lambda \right)^2},
\end{align}
where $P_t$ is the transmitted power at the transmitter.

Thus, the \gls{snr} {-- a measure of how clearly the receiver can distinguish the signal from background noise --} at the legitimate receiver, considering all impairments, can be expressed as
\begin{equation}
\text{SNR}_{\text{main}} = \frac{P_r}{N},
\end{equation}
where $N$ {is the noise power at the receiver, representing unwanted background interference picked up alongside the signal.}

\begin{figure}[t!]
    \centering
    \includegraphics[width=\linewidth]{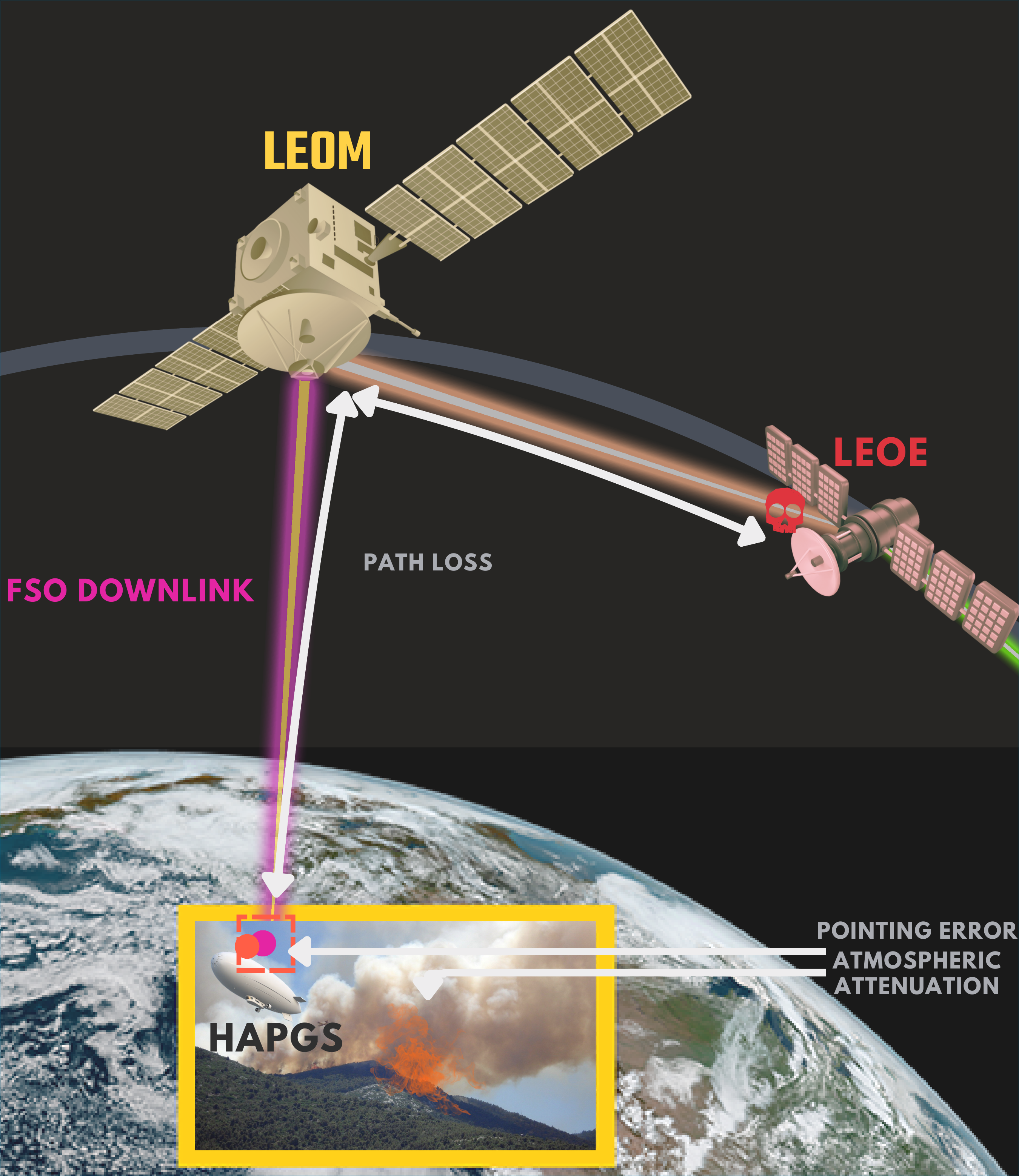}
    \caption{{Eavesdropping on the LEOM-HAPGS link.}}
    \label{fig:fso-leo=haps}
\end{figure}

\begin{figure}[t!]
    \centering
    \includegraphics[width=\linewidth]{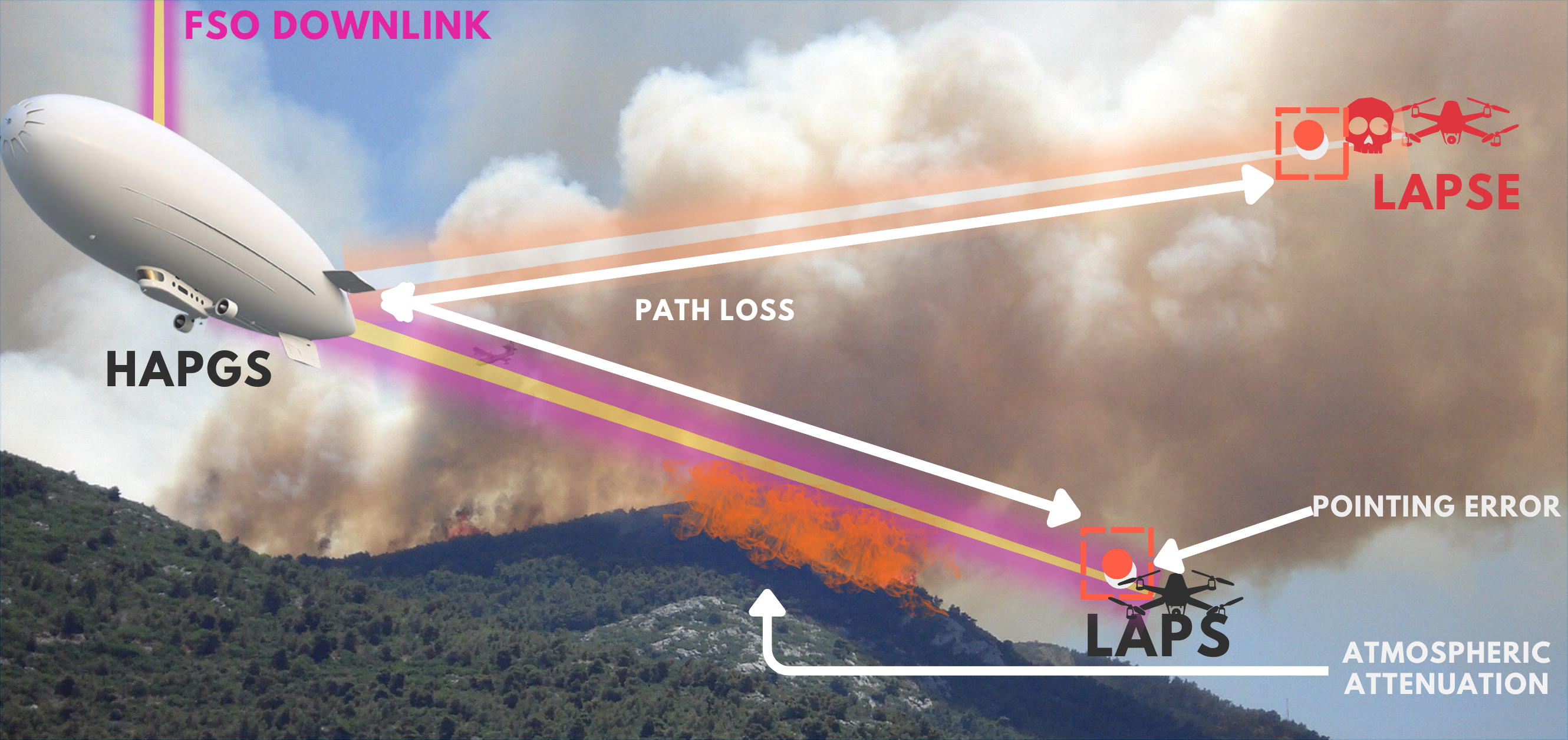}
    \caption{{Eavesdropping on the HAPGS-LAP link.}}
    \label{fig:fso=haps-laps}
\end{figure}

\subsection{Cyber Adversaries}

To model the threat technique of eavesdropping (\textit{EXF-0003}), we position a cyber adversary near, or even co-located with, one of the transmitting nodes (LEOM in the downlink, or HAPGS in the downlink) in an attempt to eavesdrop on the transmitted signal with proximity operations (\textit{REC-0005.03}). As we demonstrated before, various attenuation and errors affect the legitimate receiver’s signal reception. So if we consider that the eavesdropper experiences fewer of these impairments or is better aligned with the transmitter, its SNR may surpass that of the legitimate receiver. {We use the secrecy capacity} $C_s$, {which quantifies the maximum rate at which information can be securely transmitted without being intercepted}, defined as

\begin{equation}
C_s = \max \left( C_m - C_e, 0 \right),
\end{equation}
where $C_m$ and $C_e$ denote the capacities of the main and eavesdropper channels, respectively. These are given by

\begin{equation}
C_m = \log_2 \left(1 + \text{SNR}_{\text{main}} \right),
\end{equation}
and
\begin{equation}
C_e = \log_2 \left(1 + \text{SNR}_{\text{e}} \right),
\end{equation}

A positive secrecy capacity ensures that the legitimate receiver obtains more reliable information than the eavesdropper, thus confirming the security of the link during transmission. The opposite means that if the eavesdropper has a better SNR, the secrecy capacity will be null, implying the link is no longer secure. With that, we now present the channel model for the eavesdropper in both links.

\subsubsection{Eavesdropping on the LEOM-HAPGS Link} {LEOE} faces no impairment on-orbit and no attenuation in the stratospheric and thus is only impacted by path loss, as depicted in {Fig.}~\ref{fig:fso-leo=haps}. The received power at LEOE can be expressed as

\begin{align}
P_{r, e} &= {P_t} G_t G_e \cdot \frac{1}{\left(4\pi d_e / \lambda \right)^2}, 
\end{align}
where $P_t$ is the transmitted power at the LEOM. $G_e$ is the gain of LEOE, $d_e$ is the distance between the LEOM and LEOE nodes. Additionally, the SNR at LEOE can then be expressed as
\begin{equation}
\text{SNR}_{\text{e}} = \frac{P_{r,e}}{N_e},
\end{equation}
and $N_e$ represents the noise power at the eavesdropper.

\subsubsection{Eavesdropping on the HAPGS-LAP Link} {LAPSE} faces impairment of the atmospheric attenuation as well as the path loss, as depicted in {Fig.}~\ref{fig:fso=haps-laps}. The received power at LAPSE can be expressed as

\begin{align}
P_{r, e} &= {P_t} G_t G_e \cdot e^{-\alpha_{atm} \cdot d_e} \cdot \frac{1}{\left(4\pi d_e / \lambda \right)^2},
\end{align}
where $P_t$ is the transmitted power at the HAPGS. $G_e$ is the gain of LAPSE, $d_e$ is the distance between the HAPGS and LAPSE nodes. Additionally, the SNR at LAPSE can then be expressed as

\begin{equation}
\text{SNR}_{\text{e}} = \frac{P_{r,e}}{N_e},
\end{equation}
where $N_e$ is the noise power at the eavesdropper.

\begin{table}[t!]
\caption{Simulation Parameters}
\label{tab:simulation-params}
\centering
\renewcommand{\arraystretch}{1.6} 
\begin{tabular}{|l|l|l|}
\hline
\textbf{Notation} & \textbf{Definition}                               & \textbf{Value}    \\ \hline\hline
\multicolumn{3}{|c|}{\textbf{General Parameters}} \\ \hline
$N$                & Noise power at transmitter                       & -130 dB           \\ \hline
$N_e$              & Noise power at eavesdropper                      & -130 dB           \\ \hline
$\lambda$          & Wavelength of optical signal                     & 1550 nm           \\ \hline
\multicolumn{3}{|c|}{\textbf{Orbital Parameters}} \\ \hline
$\theta_{in, LAPS}$ & Orbit incline angle of LAPS                      & 25$^\circ$        \\ \hline
$\theta_{in, LEOM}$ & Orbit incline angle of LEOM                      & 55$^\circ$        \\ \hline
$h_{LEOM}$          & Altitude of LEOM        & 700 km            \\ \hline
$h_{LEOE}$          & Altitude of LEOM        & 600 km            \\ \hline
$h_{LAPS}$          & Altitude of LAPS   & 150 m             \\ \hline
$h_{LAPSE}$          & Altitude of LAPSE   & 1 km             \\ \hline
$h_{HAPGS}$         & Altitude of HAPGS  & 20 km             \\ \hline
\multicolumn{3}{|c|}{\textbf{Antenna Parameters}} \\ \hline
$G_t$               & Antenna gain of transmitter                      & 42.1 dB           \\ \hline
$G_r$               & Antenna gain of receiver                         & 52.1 dB           \\ \hline
\end{tabular}
\end{table}

\subsection{Secrecy Performance Analysis}
\label{quanti}

The impact of environmental stressors on secrecy capacity and trade-offs between resilience and security is analyzed for the two communication FSO links under study. This analysis is based on three assessments: \begin{enumerate*}
    \item \textbf{Impact of Atmospheric Attenuation}, which evaluates how varying atmospheric conditions affect secrecy capacity; \item \textbf{Impact of Eavesdropper Power and Distance,} which examines the influence of eavesdropper power and proximity on link security; and \item \textbf{Impact of Transmit Power}, which assesses how variations in transmit power affect secrecy capacity.
\end{enumerate*} The parameters used in each assessment, unless otherwise noted, are presented in Table \ref{tab:simulation-params}.

\subsubsection{Impact of Atmospheric Attenuation}

In the first assessment, we evaluate the secrecy capacity of the two channels versus propagation distance under varying atmospheric attenuation. The magnitude order of atmospheric attenuation for the LEOM-HAPGS link ranges from 1 dB/km to 10 dB/km whereas for the HAPGS-LAP and HAPGS-LAPSE links, the attenuation magnitude is significantly higher, ranging from 10 dB/km to 100 dB/km, reflecting the more severe effects of atmospheric disturbances. Additionally, the receiver gain is set equal to the transmitter gain, ensuring that both the transmitted and received signals are amplified to the same degree.

\paragraph{{\underline{LEOM-HAPGS and LEOM-LEOE}}}

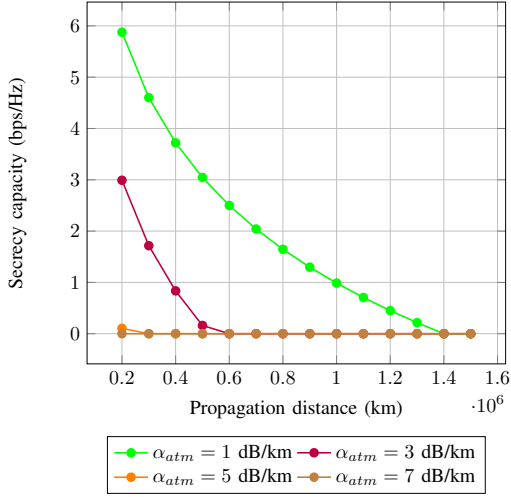
\begin{figure}[t!]
    \centering
    \scalebox{0.75}{
    \begin{tikzpicture}
    \begin{axis}[ 
        xlabel={Propagation distance (km) }, 
        ylabel={Secrecy capacity (bps/Hz)}, 
        grid=both, 
        width=9cm, 
        height=8cm, 
        title={}, 
        xtick={200000, 400000, 600000, 800000, 1000000, 1200000, 1400000, 1600000}, 
        legend style={at={(0.5,-0.2)}, anchor=north, legend columns=2}
    ]
    
    
    \addplot[mark=*, green, thick] coordinates {
        (200000, 5.87633592) (300000, 4.60147608) (400000, 3.72033089) (500000, 3.04626092)
        (600000, 2.50022393) (700000, 2.04126024) (800000, 1.64538148) (900000, 1.29732268)
        (1000000, 0.98676623) (1100000, 0.70641091) (1200000, 0.45089982) (1300000, 0.21618592)
        (1400000, 0) (1500000, 0)
    };
    \addlegendentry{$\alpha_{atm}=1$ dB/km}
    
    \addplot[mark=*, purple, thick] coordinates {
        (200000, 2.99094583) (300000, 1.71608599) (400000, 0.83494081) (500000, 0.16087084)
        (600000, 0) (700000, 0) (800000, 0) (900000, 0) (1000000, 0) (1100000, 0)
        (1200000, 0) (1300000, 0) (1400000, 0) (1500000, 0)
    };
    \addlegendentry{$\alpha_{atm}=3$ dB/km}
    
    \addplot[mark=*, orange, thick] coordinates {
        (200000, 0.10555575) (300000, 0) (400000, 0) (500000, 0) (600000, 0) (700000, 0)
        (800000, 0) (900000, 0) (1000000, 0) (1100000, 0) (1200000, 0) (1300000, 0)
        (1400000, 0) (1500000, 0)
    };
    \addlegendentry{$\alpha_{atm}=5$ dB/km}
    
    \addplot[mark=*, brown, thick] coordinates {
        (200000, 0) (300000, 0) (400000, 0) (500000, 0) (600000, 0) (700000, 0)
        (800000, 0) (900000, 0) (1000000, 0) (1100000, 0) (1200000, 0) (1300000, 0)
        (1400000, 0) (1500000, 0)
    };
    \addlegendentry{$\alpha_{atm}=7$ dB/km}
    
    
    
    
    \end{axis}
    \end{tikzpicture}}
    \caption{Secrecy capacity of LEOM - HAPGS channel versus propagation distance under varying atmospheric attenuation.}
    \label{LEOM - HAPGS-scenario1}
\end{figure}

We varied the distance of the LEOM from 200 km to 1400 km while maintaining a fixed 100 km offset for the LEOE. The HAPGS remained at a constant location, and the eavesdropper's channel was assumed to be unaffected by atmospheric attenuation. {Fig.}~\ref{LEOM - HAPGS-scenario1} shows that as the propagation distance increases, the secrecy capacity decreases. This degradation is primarily attributed to the increasing atmospheric attenuation on the main channel LEOM-HAPGS, while the eavesdropper's channel remains unperturbed. Additionally, as LEOM moves farther from HAPGS, beam misalignment and pointing errors increase, further weakening the received signal at HAPGS. In contrast, the eavesdropper's channel maintains a relatively stable performance since it is not subject to atmospheric losses or pointing errors. Moreover, while path loss remains consistent due to the fixed distance between LEOM and LEOE, the compounded effects of attenuation and pointing errors in the main channel lead to a significant drop in secrecy performance. This demonstrates that increasing the propagation distance does not necessarily improve secrecy but, in this scenario, exacerbates the vulnerability of the main channel.

\begin{figure}[tp]
    \centering
    \scalebox{0.75}{
    \begin{tikzpicture}
    
    \begin{axis}[ 
        xlabel={Propagation distance (m)}, 
        ylabel={Secrecy capacity (bps/Hz)}, 
        grid=both, 
        width=9cm, 
        height=8cm, 
        title={}, 
        legend style={at={(0.5,-0.2)}, anchor=north, legend columns=2}
    ]
    \addplot[mark=*, blue, thick] coordinates {
        (0, 1.19668049) (100, 1.37264924) (200, 1.37264924) (300, 1.19668049) 
        (400, 1.03767167) (500, 0.88369363) (600, 0.73217273) (700, 0.58211309) 
        (800, 0.43302342) (900, 0.28462490) (1000, 0.13674395) (1100, 0.00000000) 
        (1200, 0.00000000) (1300, 0.00000000) (1400, 0.00000000) (1500, 0.00000000)
    };
    \addlegendentry{$\alpha = 10$ dB/km}
    
    \addplot[mark=*, red, thick] coordinates {
        (0, 1.10000000) (100, 1.2051080) (200, 1.25051080) (300, 0.00000000)
        (400, 0.00000000) (500, 0.00000000) (600, 0.00000000) (700, 0.00000000)
        (800, 0.00000000) (900, 0.00000000) (1000, 0.00000000) (1100, 0.00000000)
        (1200, 0.00000000) (1300, 0.00000000) (1400, 0.00000000) (1500, 0.00000000)
    };
    \addlegendentry{$\alpha = 20$ dB/km}
    
    \addplot[mark=*, green, thick] coordinates {
        (0, 0.90000000) (100, 0.90781576) (200, 0.90781576) (300, 0.00000000)
        (400, 0.00000000) (500, 0.00000000) (600, 0.00000000) (700, 0.00000000)
        (800, 0.00000000) (900, 0.00000000) (1000, 0.00000000) (1100, 0.00000000)
        (1200, 0.00000000) (1300, 0.00000000) (1400, 0.00000000) (1500, 0.00000000)
    };
    \addlegendentry{$\alpha = 30$ dB/km}
    
    \addplot[mark=*, purple, thick] coordinates {
        (0, 0.00000000) (100, 0.00000000) (200, 0.00000000) (300, 0.00000000)
        (400, 0.00000000) (500, 0.00000000) (600, 0.00000000) (700, 0.00000000)
        (800, 0.00000000) (900, 0.00000000) (1000, 0.00000000) (1100, 0.00000000)
        (1200, 0.00000000) (1300, 0.00000000) (1400, 0.00000000) (1500, 0.00000000)
    };
    \addlegendentry{$\alpha = 80$ dB/km}

    \end{axis}
    \end{tikzpicture}}
    \caption{Secrecy capacity of HAPGS - LAP  channel versus propagation distance under varying atmospheric attenuation.}
    \label{HAPGS - LAPS-scenario1}
\end{figure}

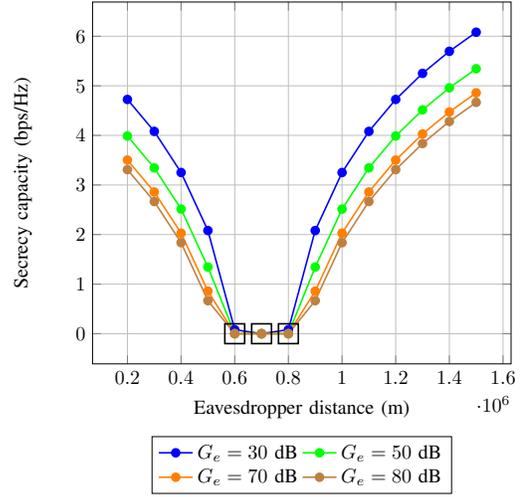
\begin{figure}[tp]
\centering
\scalebox{0.75}{
\begin{tikzpicture}
\begin{axis}[ 
    xlabel={Eavesdropper distance (m)}, 
    ylabel={Secrecy capacity (bps/Hz)}, 
    grid=both, 
    width=9cm, 
    height=8cm, 
    title={}, 
    xtick={200000, 400000, 600000, 800000, 1000000, 1200000, 1400000, 1600000}, 
    ytick={0, 1, 2, 3, 4, 5, 6}, 
    legend style={at={(0.5,-0.2)}, anchor=north, legend columns=2}
]

\addplot[mark=*, blue, thick] coordinates {
    (200000, 4.72512209) (300000, 4.0812659) (400000, 3.2511909) (500000, 2.0812659)
    (600000, 0.0812659) (700000, 0) (800000, 0.0812659) (900000, 2.0812659)
    (1000000, 3.2511909) (1100000, 4.0812659) (1200000, 4.72512209) (1300000, 5.2511909)
    (1400000, 5.69597574) (1500000, 6.0812659)
};
\addlegendentry{$G_e=30$ dB}


\addplot[mark=*, green, thick] coordinates {
    (200000, 3.9881565) (300000, 3.34430031) (400000, 2.51422531) (500000, 1.34430031)
    (600000, 0) (700000, 0) (800000, 0) (900000, 1.34430031)
    (1000000, 2.51422531) (1100000, 3.34430031) (1200000, 3.9881565) (1300000, 4.51422531)
    (1400000, 4.95901015) (1500000, 5.34430031)
};
\addlegendentry{$G_e=50$ dB}


\addplot[mark=*, orange, thick] coordinates {
    (200000, 3.50272967) (300000, 2.85887348) (400000, 2.02879848) (500000, 0.85887348)
    (600000, 0) (700000, 0) (800000, 0) (900000, 0.85887348)
    (1000000, 2.02879848) (1100000, 2.85887348) (1200000, 3.50272967) (1300000, 4.02879848)
    (1400000, 4.47358332) (1500000, 4.85887348)
};
\addlegendentry{$G_e=70$ dB}

\addplot[mark=*, brown, thick] coordinates {
    (200000, 3.31008459) (300000, 2.6662284) (400000, 1.8361534) (500000, 0.6662284)
    (600000, 0) (700000, 0) (800000, 0) (900000, 0.6662284)
    (1000000, 1.8361534) (1100000, 2.6662284) (1200000, 3.31008459) (1300000, 3.8361534)
    (1400000, 4.28093825) (1500000, 4.6662284)
};
\node (mark) [draw, black, minimum size=5pt, inner sep=5pt, thick] at (axis cs: 600000, 0) {};
\node (mark) [draw, black, minimum size=5pt, inner sep=5pt, thick] at (axis cs: 700000, 0) {};
\node (mark) [draw, black, minimum size=5pt, inner sep=5pt, thick] at (axis cs: 800000, 0) {};
\addlegendentry{$G_e=80$ dB}
\end{axis}
\end{tikzpicture}
}
\caption{{Secrecy capacity of LEOM - HAPGS channel versus varying LEOE distance under varying eavesdropper power levels.}}
\label{LEOM - HAPGS-2nd}
\end{figure}
\begin{figure}[tp]
\centering
\scalebox{0.75}{
\begin{tikzpicture}
\begin{axis}[ 
    xlabel={Eavesdropper distance (m)}, 
    ylabel={Secrecy capacity (bps/Hz)}, 
    grid=both, 
    width=9cm, 
    height=8cm, 
    title={}, 
    legend style={at={(0.5,-0.2)}, anchor=north, legend columns=2}
]
\addplot[mark=*, blue, thick] coordinates {
    (0, 3.16554336) (100, 3.15108022) (200, 3.13654422) (300, 3.12193462) 
    (400, 3.10725067) (500, 3.09249161) (600, 3.07765666) (700, 3.06274505) 
    (800, 3.04775598) (900, 3.03268864) (1000, 3.0175422) (1100, 3.00231583) 
    (1200, 2.98700868) (1300, 2.9716199) (1400, 2.9561486) (1500, 2.9405939) 
    (1600, 2.92495489)
};
\addlegendentry{$G_e = 30$ dB}


\addplot[mark=*, green, thick] coordinates {
    (0, 2.42857776) (100, 2.41411463) (200, 2.39957863) (300, 2.38496902)
    (400, 2.37028507) (500, 2.35552601) (600, 2.34069107) (700, 2.32577946)
    (800, 2.31079039) (900, 2.29572304) (1000, 2.2805766) (1100, 2.26535023)
    (1200, 2.25004309) (1300, 2.23465431) (1400, 2.21918301) (1500, 2.20362831)
    (1600, 2.1879893)
};
\addlegendentry{$G_e = 50$ dB}


\addplot[mark=*, orange, thick] coordinates {
    (0, 1.94315094) (100, 1.9286878) (200, 1.9141518) (300, 1.8995422)
    (400, 1.88485825) (500, 1.87009919) (600, 1.85526424) (700, 1.84035263)
    (800, 1.82536356) (900, 1.81029621) (1000, 1.79514977) (1100, 1.77992341)
    (1200, 1.76461626) (1300, 1.74922748) (1400, 1.73375618) (1500, 1.71820148)
    (1600, 1.70256247)
};
\addlegendentry{$G_e = 70$ dB}

\addplot[mark=*, brown, thick] coordinates {
    (0, 1.75050586) (100, 1.73604272) (200, 1.72150672) (300, 1.70689712)
    (400, 1.69221317) (500, 1.67745411) (600, 1.66261916) (700, 1.64770755)
    (800, 1.63271848) (900, 1.61765114) (1000, 1.6025047) (1100, 1.58727833)
    (1200, 1.57197118) (1300, 1.5565824) (1400, 1.5411111) (1500, 1.5255564)
    (1600, 1.50991739)
};
\addlegendentry{$G_e = 80$ dB}

\end{axis}
\end{tikzpicture}
}
\caption{Secrecy capacity of HAPGS - LAP  channel versus varying LAPSE distance under varying eavesdropper power levels.}
    \label{HAPGS - LAPS-2nd}
\end{figure}
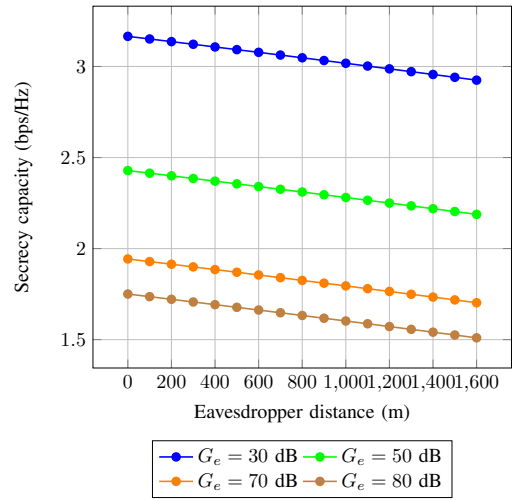

\paragraph{{\underline{HAPGS-LAP and HAPGS-LAPSE}}}

We varied the distance of the LAPS from 0 m (ground level) to 1500 m while maintaining a fixed 100 m offset for the LAPSE, with the HAPGS transmitter remaining at its set location. The atmospheric attenuation is considered for the main channel HAPGS-LAP and the eavesdropper channel HAPGS-LAPSE. {Fig.}~\ref{HAPGS - LAPS-scenario1} illustrates that as the altitude of LAP increases, the secrecy capacity decreases, particularly at higher attenuation levels. For lower attenuation values, the secrecy capacity remains relatively stable for shorter distances but rapidly diminishes as the distance increases. This is due to the compounded effects of increasing atmospheric attenuation on both the main and eavesdropper channels, as well as the misalignment on the main channel, which significantly impacts the secrecy capacity. At higher attenuation values, the main channel experiences substantial degradation, resulting in a complete loss of secrecy capacity even at shorter distances.

\subsubsection{Impact of Eavesdropper Power and Distance}

In the second assessment, we analyze the impact of eavesdropper power and distance on the secrecy capacity of the network.

\paragraph{{\underline{LEOM-HAPGS and LEOM-LEOE}}}

As shown in {Fig.}~\ref{LEOM - HAPGS-2nd}, LEOE can approach LEOM closely. As a result, secrecy capacity drops to zero at approximately 750 km, where LEOM is located, across all power levels. This indicates that no secrecy can be maintained when the eavesdropper is in close proximity to the source. {We denote this area with symbol \textbf{$\square$}.}

\paragraph{{\underline{HAPGS-LAP and HAPGS-LAPSE}}}

As shown in {Fig.}~\ref{HAPGS - LAPS-2nd}, the eavesdropper, represented by LAPSE, is constrained by altitude limitations and cannot reach the transmitted HAPGS directly. However, despite this limitation, secrecy capacity is still significantly degraded due to propagation effects and lower attenuation levels experienced at the HAPGS layer.

\noindent From this assessment, we stress that secrecy capacity is significantly influenced by both factors. When $G_e$ is low, the secrecy capacity remains relatively high, but it gradually declines as the eavesdropper moves closer and its power increases. A sharp decline is observed when $G_e$ reaches higher levels, ultimately leading to a complete loss of secrecy capacity.  Additionally, comparing the two links, we prove the impact of how the atmospheric attenuation is more pronounced in the second channel, significantly affecting both the main and eavesdropper channels compared to the first channel. This also highlights the advantage of using HAGS, as it experiences lower atmospheric losses than if it were a terrestrial station. If the ground station were positioned on the ground, it would severely degrade both links and further reduce secrecy capacity.

\begin{figure}[t!]
\centering
\scalebox{0.7}{
\begin{tikzpicture}
\begin{axis}[ 
    xlabel={Transmit power}, 
    ylabel={Capacity of main channel (bps/Hz)}, 
    ylabel near ticks,
    grid=both,
    width=9cm, 
    height=8cm, 
    title={}, 
    xtick={40, 50, 60, 70, 80, 90, 100, 110}, 
    legend style={at={(0.5,-0.2)}, anchor=north, legend columns=2}
]
\addplot[mark=*, blue, thick] coordinates {
    (40, 351.15776586) (50, 351.47969395) (60, 351.74272836) (70, 351.96512078)
    (80, 352.15776586) (90, 352.32769086) (100, 352.47969395) (110, 352.61719748)
};
\addlegendentry{$\alpha_{atm} = 1$ dB/km}

\addplot[mark=*, red, thick] coordinates {
    (40, 348.27237578) (50, 348.59430387) (60, 348.85733828) (70, 349.0797307)
    (80, 349.27237578) (90, 349.44230078) (100, 349.59430387) (110, 349.7318074)
};
\addlegendentry{$\alpha_{atm} = 3$ dB/km}

\addplot[mark=*, green, thick] coordinates {
    (40, 345.38698569) (50, 345.70891379) (60, 345.9719482) (70, 346.19434062)
    (80, 346.38698569) (90, 346.5569107) (100, 346.70891379) (110, 346.84641731)
};
\addlegendentry{$\alpha_{atm} = 5$ dB/km}

\addplot[mark=*, purple, thick] coordinates {
    (40, 342.50159561) (50, 342.82352371) (60, 343.08655811) (70, 343.30895054)
    (80, 343.50159561) (90, 343.67152061) (100, 343.82352371) (110, 343.96102723)
};
\addlegendentry{$\alpha_{atm} = 7$ dB/km}

\node (mark) [draw, purple, minimum size=5pt, inner sep=5pt, thick] at (axis cs: 40, 342.50159561) {};
\node (mark) [draw, purple, minimum size=5pt, inner sep=5pt, thick] at (axis cs: 50, 342.82352371) {};
\node (mark) [draw, purple, minimum size=5pt, inner sep=5pt, thick] at (axis cs: 60, 343.08655811) {};
\node (mark) [draw, purple, minimum size=5pt, inner sep=5pt, thick] at (axis cs: 70, 343.30895054) {};
\node (mark) [draw, purple, minimum size=5pt, inner sep=5pt, thick] at (axis cs: 80, 343.50159561) {};
\node (mark) [draw, purple, minimum size=5pt, inner sep=5pt, thick] at (axis cs: 90, 343.67152061) {};
\node (mark) [draw, purple, minimum size=5pt, inner sep=5pt, thick] at (axis cs: 100, 343.82352371) {};
\node (mark) [draw, purple, minimum size=5pt, inner sep=5pt, thick] at (axis cs: 110, 343.96102723) {};

\node (mark) [draw, green, minimum size=5pt, inner sep=5pt, thick] at (axis cs: 40, 345.38698569) {};
\node (mark) [draw, green, minimum size=5pt, inner sep=5pt, thick] at (axis cs: 50, 345.70891379) {};
\node (mark) [draw, green, minimum size=5pt, inner sep=5pt, thick] at (axis cs: 60, 345.9719482) {};
\node (mark) [draw, green, minimum size=5pt, inner sep=5pt, thick] at (axis cs: 70, 346.19434062) {};
\node (mark) [draw, green, minimum size=5pt, inner sep=5pt, thick] at (axis cs: 80, 346.38698569) {};
\node (mark) [draw, green, minimum size=5pt, inner sep=5pt, thick] at (axis cs: 90, 346.5569107) {};
\node (mark) [draw, green, minimum size=5pt, inner sep=5pt, thick] at (axis cs: 100, 346.70891379) {};
\node (mark) [draw, green, minimum size=5pt, inner sep=5pt, thick] at (axis cs: 110, 346.84641731) {};

\end{axis}
\end{tikzpicture}
}
\caption{Capacity of LEOM - HAPGS channel versus transmit power of LEOM under varying atmospheric attenuation.}
\label{LEOMHAPGSChannel}
\end{figure}
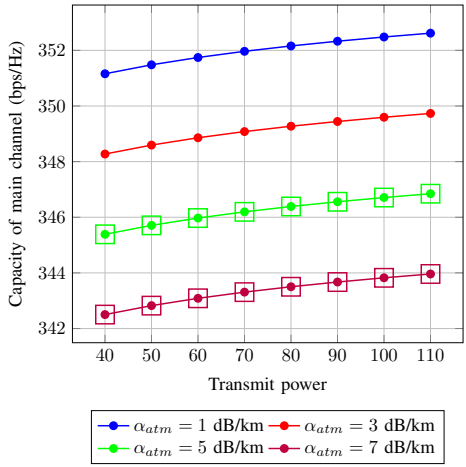

\begin{figure}[tp]
\centering
\centering
\scalebox{0.7}{
\begin{tikzpicture}

\begin{axis}[ 
    xlabel={Transmit power}, 
    ylabel={Capacity of eavesdropper channel (bps/Hz)}, 
    grid=both, 
    width=9cm, 
    height=8cm, 
    title={}, 
    xtick={40, 50, 60, 70, 80, 90, 100, 110}, 
    legend style={at={(0.5,-0.2)}, anchor=north, legend columns=2}
]
\addplot[mark=*, blue, thick] coordinates {
    (40, 345.58970651) (50, 345.9116346) (60, 346.17466901) (70, 346.39706143) 
    (80, 346.58970651) (90, 346.75963151) (100, 346.9116346) (110, 347.04913813)
};
\addlegendentry{$\alpha_{atm}=1$ dB/km}

\addplot[mark=*, red, thick] coordinates {
    (40, 345.58970651) (50, 345.9116346) (60, 346.17466901) (70, 346.39706143)
    (80, 346.58970651) (90, 346.75963151) (100, 346.9116346) (110, 347.04913813)
};
\addlegendentry{$\alpha_{atm}=3$ dB/km}

\addplot[mark=*, green, thick] coordinates {
    (40, 345.58970651) (50, 345.9116346) (60, 346.17466901) (70, 346.39706143)
    (80, 346.58970651) (90, 346.75963151) (100, 346.9116346) (110, 347.04913813)
};
\addlegendentry{$\alpha_{atm}=5$ dB/km}

\addplot[mark=*, purple, thick] coordinates {
    (40, 345.58970651) (50, 345.9116346) (60, 346.17466901) (70, 346.39706143)
    (80, 346.58970651) (90, 346.75963151) (100, 346.9116346) (110, 347.04913813)
};
\addlegendentry{$\alpha_{atm}=7$ dB/km}
\end{axis}
\end{tikzpicture}}
\caption{Capacity of LEOM - LEOE channel versus transmit power of LEOM under varying atmospheric attenuation.}
\label{LEOMLEOEChannel}
\end{figure}

\begin{figure}[tp]
\centering
\scalebox{0.7}{
\begin{tikzpicture}
\begin{axis}[ 
    xlabel={Transmit power}, 
    ylabel={Capacity of main channel (bps/Hz)}, 
    ylabel near ticks,
    grid=both,
    width=9cm, 
    height=8cm, 
    title={}, 
    xtick={40, 50, 60, 70, 80, 90, 100, 110}, 
    ytick={370.5, 372, 373, 374, 375, 376}, 
    legend style={at={(0.5,-0.2)}, anchor=north, legend columns=2}
]
\addplot[mark=*, blue, thick] coordinates {
    (40, 374.99662702) (50, 375.31855512) (60, 375.58158952) (70, 375.80398195)
    (80, 375.99662702) (90, 376.16655203) (100, 376.31855512) (110, 376.45605864)
};
\addlegendentry{$\alpha_{\text{atm}}=10$ dB/km}

\addplot[mark=*, red, thick] coordinates {
    (40, 373.55393198) (50, 373.87586008) (60, 374.13889448) (70, 374.3612869)
    (80, 374.55393198) (90, 374.72385698) (100, 374.87586008) (110, 375.0133636)
};
\addlegendentry{$\alpha_{\text{atm}}=15$ dB/km}

\addplot[mark=*, green, thick] coordinates {
    (40, 372.11123694) (50, 372.43316504) (60, 372.69619944) (70, 372.91859186)
    (80, 373.11123694) (90, 373.28116194) (100, 373.43316504) (110, 373.57066856)
};
\addlegendentry{$\alpha_{\text{atm}}=20$ dB/km}

\addplot[mark=*, orange, thick] coordinates {
    (40, 370.6685419) (50, 370.99047) (60, 371.2535044) (70, 371.47589682)
    (80, 371.6685419) (90, 371.8384669) (100, 371.99047) (110, 372.12797352)
};
\addlegendentry{$\alpha_{\text{atm}}=25$ dB/km}

\node (mark) [draw, orange, minimum size=5pt, inner sep=5pt, thick] at (axis cs: 40, 370.6685419) {};
\node (mark) [draw, orange, minimum size=5pt, inner sep=5pt, thick] at (axis cs: 50, 370.99047) {};
\node (mark) [draw, orange, minimum size=5pt, inner sep=5pt, thick] at (axis cs: 60, 371.2535044) {};
\node (mark) [draw, orange, minimum size=5pt, inner sep=5pt, thick] at (axis cs: 70, 371.47589682) {};
\node (mark) [draw, orange, minimum size=5pt, inner sep=5pt, thick] at (axis cs: 80, 371.6685419) {};
\node (mark) [draw, orange, minimum size=5pt, inner sep=5pt, thick] at (axis cs: 90, 371.8384669) {};
\node (mark) [draw, orange, minimum size=5pt, inner sep=5pt, thick] at (axis cs: 100, 371.99047) {};
\node (mark) [draw, orange, minimum size=5pt, inner sep=5pt, thick] at (axis cs: 110, 372.12797352) {};

\end{axis}
\end{tikzpicture}}
    \centering
    \caption{{Capacity of HAPGS - LAP  channel versus transmit power of HAPGS under varying atmospheric attenuation.}}
    \label{HAPGSLAPSChannel}
\end{figure}

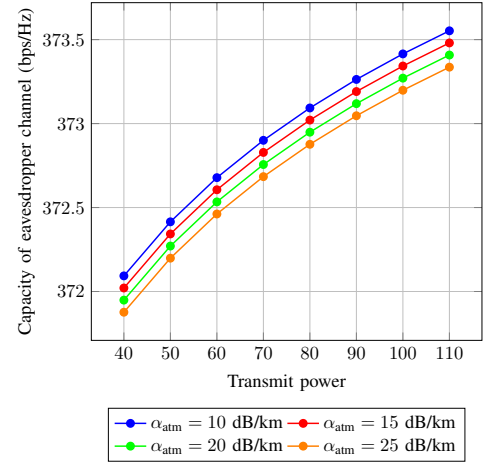
\begin{figure}[tp]
\centering
\scalebox{0.7}{
\begin{tikzpicture}
\begin{axis}[ 
    xlabel={Transmit power}, 
    ylabel={Capacity of eavesdropper channel (bps/Hz)}, 
    grid=both, 
    width=9cm, 
    height=8cm, 
    title={}, 
    xtick={40, 50, 60, 70, 80, 90, 100, 110}, 
    legend style={at={(0.5,-0.2)}, anchor=north, legend columns=2}
]
\addplot[mark=*, blue, thick] coordinates {
    (40, 372.09299651) (50, 372.41492461) (60, 372.67795901) (70, 372.90035144)
    (80, 373.09299651) (90, 373.26292152) (100, 373.41492461) (110, 373.55242813)
};
\addlegendentry{$\alpha_{\text{atm}}=10$ dB/km}

\addplot[mark=*, red, thick] coordinates {
    (40, 372.02086176) (50, 372.34278986) (60, 372.60582426) (70, 372.82821668)
    (80, 373.02086176) (90, 373.19078676) (100, 373.34278986) (110, 373.48029338)
};
\addlegendentry{$\alpha_{\text{atm}}=15$ dB/km}

\addplot[mark=*, green, thick] coordinates {
    (40, 371.94872701) (50, 372.2706551) (60, 372.53368951) (70, 372.75608193)
    (80, 372.94872701) (90, 373.11865201) (100, 373.2706551) (110, 373.40815863)
};
\addlegendentry{$\alpha_{\text{atm}}=20$ dB/km}

\addplot[mark=*, orange, thick] coordinates {
    (40, 371.87659226) (50, 372.19852035) (60, 372.46155476) (70, 372.68394718)
    (80, 372.87659226) (90, 373.04651726) (100, 373.19852035) (110, 373.33602388)
};
\addlegendentry{$\alpha_{\text{atm}}=25$ dB/km}
\end{axis}
\end{tikzpicture}}
    \centering
    \caption{Capacity of HAPGS - LAPSE channel versus transmit power of HAPGS under varying atmospheric attenuation.}
    \label{HAPGSLAPSEChannel}
\end{figure}

\subsubsection{Impact of Transmit Power}

 We examine the effect of increasing the transmitted power to counteract atmospheric losses and its impact on the secrecy of the network.

\paragraph{{\underline{LEOM-HAPGS and LEOM-LEOE}}} 

We set that LEOE is positioned at an altitude of 600 km, while LEOEM and HAPGS maintain their original position. {Fig.}~\ref{LEOMHAPGSChannel} and {Fig.}~\ref{LEOMLEOEChannel} illustrate the behavior of the channels under varying power levels. In the LEOE-LEOM channel, where atmospheric attenuation is negligible, we observe that the channel remains unaffected by changes in transmit power. However, in the LEOM-HAPGS channel, increasing attenuation leads to a significant degradation in channel performance. As the transmit power increases, the capacity of both channels improves. We denote with the symbol \textbf{$\square$} the regions where the secrecy capacity drops to zero, indicating that the eavesdropper’s channel has surpassed the main channel in capacity. This occurs because the eavesdropper is not subject to the same attenuation losses, allowing it to receive a stronger signal relative to the main channel, ultimately rendering the communication insecure.

\paragraph{{\underline{HAPGS-LAP and HAPGS-LAPSE}}}

We set that LAPSE is positioned at an altitude of 1 km, while LAPS and HAPGS maintain their original positions shown in Table \ref{tab:simulation-params}. {Fig.}~\ref{HAPGSLAPSChannel} and {Fig.}~\ref{HAPGSLAPSEChannel} illustrate the channel capacity for the HAPGS-LAP and HAPGS-LAPSE links, respectively. Due to its higher altitude, LAPSE experiences less atmospheric attenuation, resulting in a more stable channel. As the transmit power increases, channel capacity improves. However, the main channel exhibits a similar trend to the LEOM-HAPGS link, where increased attenuation leads to a substantial reduction in capacity. We also denote with the symbol  \textbf{$\square$} regions for each attenuation level where the secrecy capacity reaches zero, signifying that the eavesdropper’s channel capacity has exceeded the main channel.

\noindent This final assessment allows us to demonstrate the fundamental trade-off between increasing transmit power to counteract attenuation and the corresponding risk of enhanced eavesdropping capability, emphasizing the need for a more strategic approach to secure communication for both scenarios.

\section{System Protection Techniques}
\label{sec:protect}

The study conducted provides critical insights into how both environmental factors and adversarial actions can compromise the confidentiality and integrity of the links. By analyzing the impact of these influences, we identified vulnerabilities that may threaten the secure transmission. The key takeaways from this study underscore several important challenges:

\begin{itemize}
    \item Atmospheric attenuation, especially in short-range communication HAPGS-LAP and long-range communication LEOM-HAPGS, notably reduces the secrecy capacity of the communication link. This degradation is more pronounced at lower altitudes due to higher attenuation in short-range links, and increases with longer propagation distances in the long-range link, while the eavesdropper's channel remains unaffected by these environmental factors.
    \item As the LEOM moves farther from the HAPGS, the compounded effects of beam misalignment and pointing errors further degrade the communication quality, making it easier for eavesdroppers to intercept signals.
    \item The altitude of the eavesdropper's position also plays a role in affecting secrecy; for instance, higher altitudes lead to less atmospheric attenuation, thereby enhancing eavesdropping capabilities when the LEOM is transmitting either to HAPGS (as demonstrated) and also to the rest of the constellations.
    \item Increasing transmit power to compensate for attenuation risks improving eavesdropper's reception, leading to the potential loss of secrecy capacity if not carefully managed.
\end{itemize}

\subsection{{Protection Techniques}}

{Building on this, we define protection techniques with explicit traceability to the degradation factors observed. Two additional technical functions are introduced to directly address the interplay between environmental effects and security exposure:}

\begin{itemize}

    \item \textit{{Atmospheric Attenuation Compensation} (\textit{{PRO-001}}}): {Adjusts communication parameters to maintain link performance despite signal degradation caused by environmental disturbances. }

    \item \textit{{Beam Alignment and Tracking} (\textit{{PRO-002}}}): {Ensures communication stability by automatically adjusting the beam direction to compensate for environmental factors. }
\end{itemize}    



\vspace{0.5cm}

\noindent \textbf{\textsc{Subprocess 3: Establish protection techniques to technical traceability}}

\noindent {This subprocess focuses on consolidating all protection techniques needed to ensure confidentiality, integrity, and resilience. We leverage the SPARTA \cite{aerospace2023sparta} cyber protection mappings to anchor standard countermeasures, such as:}

\begin{itemize}
    \item \textit{TRANSEC} (\textit{CM0029}): Protects against unauthorized interception and ensures secure communication even under adverse environmental conditions that lead to signal degradation.

    \item \textit{COMSEC} (\textit{CM0002}): Safeguards the integrity and confidentiality of data during transmission, preventing attacks like replay attacks or signal jamming, and ensuring secure communications even in adverse conditions.

    \item \textit{TEMPEST} (\textit{CM0003}): Mitigates risks of signal interception by securing the electromagnetic emissions from FSO/RF links, especially under environmental factors that could unintentionally leak signals.

    \item \textit{Secure Command Mode(s)}: (\textit{CM0055}) Protects against unauthorized command injections and ensures communication between HAPGS and LAP or LEOM and HAPGS remains secure under harsh environmental conditions. 

\end{itemize}    

{The two newly introduced techniques—(PRO-001, PRO-002) —extend the SPARTA baseline by addressing system-level degradations that traditional cyber mappings do not fully capture. }

\vspace{0.5cm}

\noindent \textbf{\textsc{Subprocess 4: Establish Secure Blocks}}

\noindent In the final subprocess, we focus on defining secure blocks, which encapsulate the functional technical requirements that \textsc{SHALL} be implemented to deliver the protection techniques for the identified system boundary. 

\subsection{{Secure Blocks}}

{We articulate the technical security requirements for the identified system boundary. The following are exemplary \textsc{SHALL} statements corresponding to each technical element.}

\subsubsection{{FSO/RF Management}}

\begin{quote}
\textit{The FSO/RF Management \textsc{SHALL} ensure that atmospheric attenuation compensation mechanisms are in place to adjust transmission power dynamically in response to environmental conditions, such as smoke or extreme heat, to mitigate signal degradation and maintain reliable communication.
}
\end{quote}

\begin{quote}
\textit{The FSO/RF Management \textsc{SHALL} ensure that beam alignment and tracking are in place to correct for potential beam misalignment caused by environmental factors to ensure uninterrupted communication.
}
\end{quote}

\subsubsection{{Space-to-X Links}}

\begin{quote}
\textit{  
    The  Space-to-X Links \textsc{SHALL}  ensure that Secure Command Modes (CM0055) are in place for communication with spacecraft to restrict command acceptance based on geographic location, operational modes, or time windows, to prevent unauthorized commands during periods of system vulnerability due to environmental factors.}
\end{quote}

\begin{quote}
\textit{  
    The  Space-to-X Links \textsc{SHALL} employ TRANSEC (CM0029) measures to secure the communication link between LEOM and HAPGS, ensuring that the channel is resistant to jamming, spoofing, and eavesdropping, even in degraded environmental conditions.}
\end{quote}

\definecolor{temColor}{RGB}{128,0,128} 
\definecolor{tranColor}{RGB}{0,128,128} 
\definecolor{comColor}{RGB}{0,100,0} 
\definecolor{cmdColor}{RGB}{204,102,0} 
\definecolor{envColor}{RGB}{150,75,0} 
\definecolor{envColor2}{RGB}{120,15,0} 

\section{{Key Takeaways and Discussion}}
\label{takeways}
{
The presented system and subsequent analysis and investigation have led to some key takeaways. First, the identification and risk analysis for implicit attack vectors enabled by the physical world, e.g., communication disturbances due to natural events impairing link confidentiality, are important and deserve more attention. The dual nature of adversaries (intentional and non-intentional) impacting security and resilience should be considered in hybrid models integrating physical and cyber worlds. Second, the current landscape of cyber-physical systems serving use cases such as wildfire monitoring is complex, with various adversarial actors. More comprehensive yet tractable models and defensive techniques are crucial.  For the former, as seen in our component-driven analysis detailed in Appendix~\ref{appendix1}, complexity management for the security analysis and protection strategies is fundamental. For the latter, the heterogeneous and distributed nature of these systems makes it challenging to implement and enforce defense and mitigation techniques. In that regard, a \textit{secure-by-design} paradigm will alleviate this challenge since the system components are designed and integrated with security, not as an afterthought, as demonstrated by the \textit{IEEE P3536 Standard for Space System Cybersecurity Design} \cite{IEEE2025P3536} process. Third, physical disturbances have a major impact in addition to the cyber adversaries, as seen in our numerical analysis. Nevertheless, a more comprehensive analysis is needed, compared to ours, which is solely focused on specific link types.}

{Based on these key takeaways, we can enumerate some limitations in our system: Our analysis in Section~\ref{case-study} considers physical impairments in the wireless links. However, that contribution could be extended to other segments and higher layers in the network stack, as well as cascaded scenarios such as a disaster causing power outages, which in turn leads to loss of some network capabilities, degrading the security level. Moreover, the analysis could be dynamic and even event-driven based on emerging new threats. Essentially, it can be recurrent to have an adaptive nature. In our work, we have provided our contribution based on a snapshot of the envisaged scenario, employed HetNet architecture, and the physical/environmental circumstances. The considered HetNet and system model could also face challenges from the deployment and practical aspects: First, spectrum management is not trivial when you have these different wireless nodes, such as HAPS and UAVs, operating in emergency scenarios in an ad hoc manner. Moreover, air traffic management, assuring safety and smooth operation, is challenging since the infrastructure could be damaged, and communications are also impaired, as described in our analysis. The two aspects are linked to the regulatory~\cite{8350752} and techno-economic aspects~\cite{10746260}, as well.}

{There are also energy constraints~\cite{9424181} in energy-limited and wireless ad hoc heterogeneous networks, including aerial platforms such as HAPS considered in our scenario. The energy and computational constraints of such platforms also have an impact on the performance of wildfire management since they will limit the sensing, communication, and computational capabilities of network nodes. From the security perspective, these limitations may also impair the security assurance in our system since protection techniques discussed in Section~\ref{sec:protect} and security-related system functions such as monitoring, encryption, and attack detection also require computational and communication resources.}

{The identified key takeaways and limitations also highlight some potential future work towards secure and robust HetNets for wildfire monitoring scenarios. An apparent research direction is to extend our link-focused numerical analysis in Section~\ref{case-study} to an end-to-end scenario with different system segments/domains. This calls for a hybrid analysis that integrates the environmental factors and adversarial cyber threats for performance analysis in a comprehensive use case, including all system segments. However, this more holistic view should be accompanied by a complexity analysis since the envisaged secure systems engineering methodology could suffer from scalability and resource challenges in such scenarios. Another important research direction is to integrate potential on-orbit disturbances in space and new link technologies to address the space segment cybersecurity and resilience. Additionally, future work should focus on extending the current channel modeling framework to account for real-time dynamics observed during active wildfire scenarios. This includes modeling rapidly changing smoke density, particle composition, and visibility degradation, which significantly affect atmospheric attenuation. Incorporating such adaptive models that respond to sensor-derived environmental inputs would allow the system to maintain performance under high-load, time-critical conditions. }

\begin{table*}[t]
    \centering
    \renewcommand{\arraystretch}{1.5}
    \caption{{Full view of threats techniques and protection techniques mapped to system boundary.}}
    \begin{tabular}{|p{2.6cm}|l|p{9cm}|}
    \hline
        \textbf{Targeted Technical Element} & \textbf{Threat Technique} & \textbf{Protection Technique} \\        \hline\hline
        \multicolumn{3}{|c|}{\cellcolor{gray!20}\textbf{Environmental Adversaries}} \\  \hline
        \rowcolor{gray!15}
         &  Atmospheric Signal Attenuation (\textit{NAT-001}) & \textcolor{envColor2}{Atmospheric Attenuation Compensation (\textit{PRO-001})} \\
        \rowcolor{gray!15}
        \multirow{1}{*}{\cellcolor{gray!15}FSO/RF Management}  & Beam Misalignment (\textit{NAT-002}) & \textcolor{envColor}{Beam Alignment and Tracking (\textit{PRO-002})} \\
        \rowcolor{gray!15}
         &  Increased Background Noise (\textit{NAT-003}) & \textcolor{envColor}{Beam Alignment and Tracking (\textit{PRO-002})} \\
        \hline\hline
        \multicolumn{3}{|c|}{\cellcolor{blue!10} \textbf{Cyber Adversaries}} \\         \hline
        \rowcolor{blue!10}
           & Eavesdropping (\textit{EXF-0003}) & \textcolor{temColor}{TEMPEST (\textit{CM0003})}, \textcolor{tranColor}{TRANSEC (\textit{CM0029})}, \textcolor{comColor}{COMSEC (\textit{CM0002})} \\
        \rowcolor{blue!10}
         & Uplink Intercept (\textit{REC-0005.01}) & \textcolor{tranColor}{TRANSEC (\textit{CM0029})}, \textcolor{comColor}{COMSEC (\textit{CM0002})} \\
        \rowcolor{blue!10}
         & Downlink Intercept (\textit{REC-0005.02}) & \textcolor{tranColor}{TRANSEC (\textit{CM0029})}, \textcolor{comColor}{COMSEC (\textit{CM0002})} \\
        \rowcolor{blue!10}
           & Proximity Operations (\textit{REC-0005.03}) & \textcolor{temColor}{TEMPEST (\textit{CM0003})}, \textcolor{tranColor}{TRANSEC (\textit{CM0029})}, \textcolor{comColor}{COMSEC (\textit{CM0002})} \\
        \rowcolor{blue!10}
          & Active Scanning RF/Optical (\textit{REC-0005.04}) & \textcolor{tranColor}{TRANSEC (\textit{CM0029})}, \textcolor{comColor}{COMSEC (\textit{CM0002})} \\
        \rowcolor{blue!10}
        \multirow{1}{*}{\cellcolor{blue!10} Space-to-X Links} & Crosslink via Compromised Neighbor (\textit{IA-0003}) & \textcolor{cmdColor}{Secure Command Mode(s) (\textit{CM0055})}, \textcolor{temColor}{TEMPEST (\textit{CM0003})}, \textcolor{comColor}{COMSEC (\textit{CM0002})}, \textcolor{tranColor}{TRANSEC (\textit{CM0029})} \\
        \rowcolor{blue!10}
         & Replay Attack (\textit{EX-0001}) & \textcolor{cmdColor}{Secure Command Mode(s) (\textit{CM0055})}, \textcolor{tranColor}{TRANSEC (\textit{CM0029})}, \textcolor{comColor}{COMSEC (\textit{CM0002})} \\
        \rowcolor{blue!10}
         & Prevent Downlink (\textit{DE-0002}) & \textcolor{cmdColor}{Alternate Communications Paths (\textit{CM0070})}, \textcolor{comColor}{COMSEC (\textit{CM0002})}, \textcolor{tranColor}{TRANSEC (\textit{CM0029})} \\
        \rowcolor{blue!10}
         & Jam Link Signal (\textit{DE-0002.02}) & \textcolor{cmdColor}{Alternate Communications Paths (\textit{CM0070})}, \textcolor{comColor}{COMSEC (\textit{CM0002})}, \textcolor{tranColor}{TRANSEC (\textit{CM0029})} \\
        \rowcolor{blue!10}
         & Constellation Hopping via Crosslink (\textit{LM-0003}) & \textcolor{comColor}{COMSEC (\textit{CM0002})}, \textcolor{temColor}{TEMPEST (\textit{CM0003})}, \textcolor{tranColor}{TRANSEC (\textit{CM0029})} \\
        \hline
    \end{tabular}
    \label{tab:full view}
\end{table*}

\section{Conclusion} \label{conclusion}

In this paper, we analyze the impact of both environmental disruptions and cyber adversarial threats on communication links within a HetNet wildfire management network, focusing on FSO/RF management and space-to-X links. Through detailed modeling and simulations of an eavesdropper, we examine the effects of atmospheric attenuation and other environmental disturbances on the secrecy capacity of two communication links. Additionally, we demonstrate how the system should be designed in alignment with the IEEE~P3536 to help avoid mission failure in the face of the two adversaries' domains. Future work will focus on integrating the impact of on-orbit conditions and expanding the analysis to include all segments and different links of the network.

\appendices
\section{Full View of System Design}
\label{appendix1}

{The full view of threat techniques and protection techniques mapped to the system boundary is provided in Table~\ref{tab:full view}} where we categorize adversaries into cyber and environmental domains.

\section*{Acknowledgment}
This work was supported in part by the Tier 1 Canada Research Chair Program, the Natural Sciences and Engineering Research Council of Canada (NSERC) Discovery Program, and the U.S. Department of Defense, Office of the Under Secretary of Defense for Research and Engineering (OUSD(R\&E)).

\bibliographystyle{IEEEtran}
\bibliography{citations}

\begin{thebibliography}{10}
\providecommand{\url}[1]{#1}
\csname url@samestyle\endcsname
\providecommand{\newblock}{\relax}
\providecommand{\bibinfo}[2]{#2}
\providecommand{\BIBentrySTDinterwordspacing}{\spaceskip=0pt\relax}
\providecommand{\BIBentryALTinterwordstretchfactor}{4}
\providecommand{\BIBentryALTinterwordspacing}{\spaceskip=\fontdimen2\font plus
\BIBentryALTinterwordstretchfactor\fontdimen3\font minus \fontdimen4\font\relax}
\providecommand{\BIBforeignlanguage}[2]{{%
\expandafter\ifx\csname l@#1\endcsname\relax
\typeout{** WARNING: IEEEtran.bst: No hyphenation pattern has been}%
\typeout{** loaded for the language `#1'. Using the pattern for}%
\typeout{** the default language instead.}%
\else
\language=\csname l@#1\endcsname
\fi
#2}}
\providecommand{\BIBdecl}{\relax}
\BIBdecl

\bibitem{owid-wildfires}
\BIBentryALTinterwordspacing
V.~Samborska and H.~Ritchie, ``Wildfires,'' \emph{Our World in Data}, 2024, accessed: 21 March 2025. [Online]. Available: \url{https://ourworldindata.org/wildfires}
\BIBentrySTDinterwordspacing

\bibitem{NSW2021Fire}
L.~Rumpff, S.~M. Legge, S.~van Leeuwen, B.~A. Wintle, and J.~C.~Z. Woinarski, Eds., \emph{{Australia{\textquoteright}s Megafires: Biodiversity Impacts and Lessons from 2019–2020}}.\hskip 1em plus 0.5em minus 0.4em\relax CSIRO Publishing, 2023.

\bibitem{Danielle2025LAWildfires}
\BIBentryALTinterwordspacing
M.~Danielle, ``{AccuWeather Estimates More Than \$250 Billion in Damages and Economic Loss from LA Wildfires},'' \emph{AccuWeather}, January 2025, accessed: 21 March 2025. [Online]. Available: \url{https://www.accuweather.com}
\BIBentrySTDinterwordspacing

\bibitem{doi:10.1073/pnas.1617394114}
J.~K. Balch, B.~A. Bradley, J.~T. Abatzoglou, R.~C. Nagy, E.~J. Fusco, and A.~L. Mahood, ``{Human-started wildfires expand the fire niche across the United States},'' \emph{Proceedings of the National Academy of Sciences}, vol. 114, no.~11, pp. 2946--2951, 2017.

\bibitem{Campbell2021IntentionalFires}
\BIBentryALTinterwordspacing
R.~Campbell, ``Intentional structure fires,'' National Fire Protection Association (NFPA), Tech. Rep., September 2021, accessed: 21 March 2025. [Online]. Available: \url{https://www.nfpa.org/education-and-research/research/nfpa-research/fire-statistical-reports/intentional-fires}
\BIBentrySTDinterwordspacing

\bibitem{Haynes_2019}
K.~Haynes, K.~Short, G.~Xanthopoulos, D.~Viegas, L.~M. Ribeiro, and R.~Blanchi, \emph{Wildfires and WUI Fire Fatalities}.\hskip 1em plus 0.5em minus 0.4em\relax Springer International Publishing, 2019, p. 1–16.

\bibitem{Blumenfeld2020Wildfires}
\BIBentryALTinterwordspacing
J.~Blumenfeld, ``{Wildfires Can't Hide from Earth Observing Satellites},'' \emph{NASA Earthdata}, November 2020, accessed: 21 March 2025. [Online]. Available: \url{https://www.earthdata.nasa.gov/news/feature-articles/wildfires-cant-hide-from-earth-observing-satellites}
\BIBentrySTDinterwordspacing

\bibitem{KopardekarGrindle2021}
P.~Kopardekar and L.~Grindle, ``{NASA ARMD Wildfire Management Workshop},'' in \emph{NASA ARMD Wildfire Management Workshop}.\hskip 1em plus 0.5em minus 0.4em\relax NASA Aeronautics Research Institute (NARI), 2021.

\bibitem{McHugh-Johnson2025FireSat}
\BIBentryALTinterwordspacing
M.~McHugh-Johnson, ``{Inside the Launch of FireSat, a System to Find Wildfires Earlier},'' \emph{Google Blog}, March 2025, accessed: 21 March 2025. [Online]. Available: \url{https://blog.google/technology/ai/inside-firesat-launch-muon-space/}
\BIBentrySTDinterwordspacing

\bibitem{CSA2025WildFireSat}
\BIBentryALTinterwordspacing
{Canadian Space Agency}, ``Wildfiresat,'' February 2025, accessed: 21 March 2025. [Online]. Available: \url{https://www.asc-csa.gc.ca/eng/satellites/wildfiresat/}
\BIBentrySTDinterwordspacing

\bibitem{6GWorld2025HAPS}
\BIBentryALTinterwordspacing
{6GWorld}, ``{Will HAPS Become a Tool to Fight Wildfires? Experts Think So},'' \emph{6GWorld}, 2025, accessed: 21 March 2025. [Online]. Available: \url{https://www.6gworld.com/exclusives/will-haps-become-a-tool-to-fight-wildfires-experts-think-so/}
\BIBentrySTDinterwordspacing

\bibitem{Burns2023HAPS}
A.~J. Burns, M.~Johnson, J.~Jung, and M.~Fladeland, ``{High Altitude Platform System (HAPS) Communication Support for Wildland Firefighting: Annual Estimates and Considerations},'' NASA Ames Research Center, Tech. Rep. NASA/TM–20230018267, December 2023.

\bibitem{10589561}
G.~K. Pandey, D.~S. Gurjar, S.~Yadav, Y.~Jiang, and C.~Yuen, ``{UAV-Assisted Communications With RF Energy Harvesting: A Comprehensive Survey},'' \emph{IEEE Communications Surveys \& Tutorials}, vol.~27, no.~2, pp. 782--838, 2025.

\bibitem{8030545}
B.~Feng, H.~Zhou, H.~Zhang, G.~Li, H.~Li, S.~Yu, and H.-C. Chao, ``{HetNet: A Flexible Architecture for Heterogeneous Satellite-Terrestrial Networks},'' \emph{IEEE Network}, vol.~31, no.~6, pp. 86--92, 2017.

\bibitem{9129403}
R.~Swaminathan, S.~Sharma, and A.~S. MadhuKumar, ``{Performance Analysis of HAPS-Based Relaying for Hybrid FSO/RF Downlink Satellite Communication},'' in \emph{IEEE 91st Vehicular Technology Conference (VTC2020-Spring)}, 2020, pp. 1--5.

\bibitem{10570308}
C.~T. Nguyen, Y.~M. Saputra, N.~V. Huynh, T.~N. Nguyen, D.~T. Hoang, D.~N. Nguyen, V.-Q. Pham, M.~Voznak, S.~Chatzinotas, and D.-H. Tran, ``{Emerging Technologies for 6G Non-Terrestrial-Networks: From Academia to Industrial Applications},'' \emph{IEEE Open Journal of the Communications Society}, vol.~5, pp. 3852--3885, 2024.

\bibitem{9655260}
O.~Ben~Yahia, E.~Erdogan, G.~Karabulut~Kurt, I.~Altunbas, and H.~Yanikomeroglu, ``{A Weather-Dependent Hybrid RF/FSO Satellite Communication for Improved Power Efficiency},'' \emph{IEEE Wireless Communications Letters}, vol.~11, no.~3, pp. 573--577, 2022.

\bibitem{9832657}
M.~Matracia, N.~Saeed, M.~A. Kishk, and M.-S. Alouini, ``{Post-Disaster Communications: Enabling Technologies, Architectures, and Open Challenges},'' \emph{IEEE Open Journal of the Communications Society}, vol.~3, pp. 1177--1205, 2022.

\bibitem{10185601}
A.~H. Arani, P.~Hu, and Y.~Zhu, ``{HAPS-UAV-Enabled Heterogeneous Networks: A Deep Reinforcement Learning Approach},'' \emph{IEEE Open Journal of the Communications Society}, vol.~4, pp. 1745--1760, 2023.

\bibitem{10850060}
O.~Ben~Yahia, W.~Ferguson, S.~Chakravarty, N.~Benchoubane, G.~Karabulut~Kurt, G.~Gür, and G.~Falco, ``Securing satellite link segment: A secure-by-component design,'' in \emph{IEEE International Conference on Wireless for Space and Extreme Environments (WiSEE)}, 2024, pp. 177--182.

\bibitem{10539344}
M.~Elamassie and M.~Uysal, ``{FSO}-based multi-layer airborne backhaul networks,'' \emph{IEEE Transactions on Vehicular Technology}, vol.~73, no.~10, pp. 15\,004--15\,019, 2024.

\bibitem{10615868}
P.~G. Madoery, J.~A. Fraire, J.~M. Finochietto, H.~Yanikomeroglu, and G.~Karabulut~Kurt, ``{A Novel Non-Terrestrial Networks Architecture: All Optical LEO Constellations with High-Altitude Ground Stations},'' in \emph{IEEE International Conference on Communications Workshops (ICC Workshops)}, 2024, pp. 1115--1120.

\bibitem{doi:10.2514/6.2023-4800}
N.~Boschetti, J.~Slay, J.~Plotnek, G.~Karabulut~Kurt, and G.~Falco, ``{Atmospheric Outposts Security: Attack Tree Analysis of High Altitude Air Platforms (HAPSs)},'' in \emph{ASCEND 2023}.\hskip 1em plus 0.5em minus 0.4em\relax AIAA, October 2023.

\bibitem{MOHSAN2023100697}
S.~A.~H. Mohsan, M.~A. Khan, and H.~Amjad, ``{Hybrid FSO/RF networks: A review of practical constraints, applications and challenges},'' \emph{Opt. Switch. Netw.}, vol.~47, no.~C, Feb. 2023.

\bibitem{9691902}
Y.~Wang, Y.~Tong, and Z.~Zhan, ``{On Secrecy Performance of Mixed RF-FSO Systems With a Wireless-Powered Friendly Jammer},'' \emph{IEEE Photonics Journal}, vol.~14, no.~2, pp. 1--8, 2022.

\bibitem{9806158}
O.~Ben~Yahia, E.~Erdogan, and G.~Karabulut~Kurt, ``{HAPS-Assisted Hybrid RF-FSO Multicast Communications: Error and Outage Analysis},'' \emph{IEEE Transactions on Aerospace and Electronic Systems}, vol.~59, no.~1, pp. 140--152, 2023.

\bibitem{9238951}
Y.~Shi, Y.~Gao, and Y.~Xia, ``{Secrecy Performance Analysis in Internet of Satellites: Physical Layer Security Perspective},'' in \emph{IEEE/CIC International Conference on Communications in China (ICCC)}, 2020, pp. 1185--1189.

\bibitem{IEEE2025P3536}
\BIBentryALTinterwordspacing
{IEEE Standards Association}, \emph{Standard for Space System Cybersecurity Design}, Std. IEEE P3536, 2025, accessed: 21 March 2025. [Online]. Available: \url{https://standards.ieee.org/ieee/3536/11916/}
\BIBentrySTDinterwordspacing

\bibitem{10592289}
A.~Viswanathan, B.~Bailey, K.~Tan, and G.~Falco, ``Secure-by-component: A system-of-systems design paradigm for securing space missions,'' in \emph{Security for Space Systems (3S)}, 2024, pp. 1--9.

\bibitem{10795027}
G.~Falco, N.~Boschetti, A.~Viswanathan, B.~Bailey, C.~Maple, G.~Karabulut~Kurt, J.~Willbold, J.~Slay, E.~Birrane, D.~Logsdon, S.~Bennett, W.~Ferguson, J.~Curbo, J.~Oakley, M.~Schloegel, S.~Hagen, J.~Sigholm, C.~Mehlman, R.~Thummala, M.~Calabrese, Y.~Shah, A.~T. Le, K.~Tan, E.~Miller, G.~Epiphaniou, U.~I. Atmaca, W.~C. Henry, G.~Gür, R.~V. Segate, and O.~Ben~Yahia, ``Minimum requirements for space system cybersecurity - ensuring cyber access to space,'' in \emph{IEEE 10th International Conference on Space Mission Challenges for Information Technology (SMC-IT)}, 2024, pp. 78--88.

\bibitem{aerospace2023sparta}
\BIBentryALTinterwordspacing
{The Aerospace Corporation}, ``{SPARTA: Space Attack Research and Tactic Analysis},'' 2023, [Accessed: 01.10.2024]. [Online]. Available: \url{https://aerospace.org/sparta}
\BIBentrySTDinterwordspacing

\bibitem{Kim2001ComparisonOL}
I.~I. Kim, B.~McArthur, and E.~J. Korevaar, ``{Comparison of laser beam propagation at 785 nm and 1550 nm in fog and haze for optical wireless communications},'' in \emph{SPIE Optics East}, 2001.

\bibitem{Green2019}
J.~L. Green, B.~W. Welch, and R.~M. Manning, ``{Optical Communication Link: Atmospheric Attenuation Model},'' Glenn Research Center, Cleveland, Ohio, Tech. Rep., February 2019.

\bibitem{7000804}
M.~Ijaz, Z.~Ghassemlooy, A.~Gholami, and X.~Tang, ``{Smoke attenuation in free space optical communication under laboratory controlled conditions},'' in \emph{International Symposium on Telecommunications (IST'2014)}, 2014, pp. 758--762.

\bibitem{8350752}
D.~Yuniarti, ``{Regulatory challenges of broadband communication services from High Altitude Platforms (HAPs)},'' in \emph{International Conference on Information and Communications Technology (ICOIACT)}, 2018, pp. 919--922.

\bibitem{10746260}
L.~Toka, M.~Konrad, A.~Pekar, and G.~Biczók, ``Integrating the skies for {6G}: Techno-economic considerations of {LEO}, {HAPS}, and {UAV} technologies,'' \emph{IEEE Communications Magazine}, vol.~62, no.~11, pp. 44--51, 2024.

\bibitem{9424181}
O.~M. Bushnaq, A.~Chaaban, and T.~Y. Al-Naffouri, ``{The Role of UAV-IoT Networks in Future Wildfire Detection},'' \emph{IEEE Internet of Things Journal}, vol.~8, no.~23, pp. 16\,984--16\,999, 2021.

\end{thebibliography}

\end{document}